# Omni-Photonic Base Station: A Three-Functional-Domain Photonic Architecture for Evolutionary 6G Wireless Infrastructure

Dapeng Wang[1]， Xiaoxiong Song[1], Ziling Fu[1], Wenyang Cheng[1], Jiayao Wang[1],Xiaogang Yan[1], Ze Wang[1], Min Zhang[1], Jingdi Liu[2], Nan Li[1*]

[1]*China Mobile Research Institute, 32 Xuanwumen West Street, Xicheng District, Beijing, 100053, China.*
[2]*China Mobile Communications Group Co., Ltd., 29 Finance Street, Xicheng District, Beijing 100032, China.*

**Corresponding author:* linan@chinamobile.com

## Abstract

6G mobile communications impose immersive communication demands for higher data rates and massive connectivity, while emerging integrated sensing–computing–intelligence scenarios require base stations to concurrently enhance computing capability, guarantee service latency, and realize sensing. We propose the Omni-Photonic Base Station (Omni-PBS)—a progressive evolutionary architecture that introduces photonic technologies into three functional domains, namely baseband processing, fronthaul transmission, and the RF front-end, and obtains system-level gains through cross-domain co-design. Traditional electronic base stations can hardly satisfy the microsecond-level latency demanded by on-site AI inference under power constraints; digital fronthaul faces an explosive growth in transmission bandwidth brought by large-bandwidth communications; and electronic RF front-ends encounter device-level physical bottlenecks in millimeter-wave and terahertz bands. By introducing photonics, Omni-PBS harnesses the inherent physical advantages of ultra-broad bandwidth, ultra-low propagation latency, and native parallelism to transcend the aforementioned electronic bottlenecks and fulfill the compound demands of 6G scenarios.

The optical baseband computing domain employs photonic accelerators to perform linear computation-intensive tasks, improving the energy efficiency of AI inference by one to two orders of magnitude. The analog optical fronthaul domain replaces digital fronthaul with analog radio-over-fiber (A-RoF), which eliminates the ADCs/DACs and digital intermediate-frequency (IF) chips in the remote unit and substantially reduces fronthaul bandwidth requirements. The microwave-photonic RF domain breaks through the bandwidth and frequency limitations of electronic RF front-ends via optical true-time-delay (TTD) beamforming, programmable photonic filtering, and optical heterodyne frequency conversion. The three domains yield system-level gains beyond single-domain summation through four co-design principles: end-to-end optical-domain continuity (maintaining the optical signal path throughout the chain), cross-domain co-design, optical computing resource scheduling, and joint optimization of functional splitting. This paper presents the concept, scheme, and collaborative design of the Omni-PBS, analyzes engineering constraints such as optical computing precision, analog link linearization, and photonic RF integration density, and discusses open problems and future directions.

## 1 Introduction

The evolution of mobile communications from 5G to 6G is undergoing the most profound paradigm shift in base station architecture since the introduction of C-RAN. 6G raises immersive communication demands for higher data rates and massive connectivity, while new integrated sensing–computing–intelligence scenarios require base stations to concurrently enhance computing capability, guarantee service latency, and realize sensing. In Rel-18/19, 3GPP launched studies on integrated sensing and communication, and Rel-20 further elevates AI/ML from an optional enhancement to a native air-interface capability; the ultra-massive antenna arrays (256 channels and 1024 or more antenna elements) and instantaneous bandwidth of 400 MHz to 1 GHz that 6G is expected to support impose engineering requirements on the fronthaul link and RF front-end far exceeding those of 5G[1]. Integrated sensing and communication requires the RF front-end to process both communication and sensing signals simultaneously; the AI-native air interface requires the baseband to execute neural-network inference in real time; and immersive communications with high data rates and massive connectivity require large-scale distributed remote deployment. Together, these force traditional electronic base stations into simultaneous bottlenecks along three dimensions: energy-efficient low-latency computing, large-bandwidth fronthaul transmission, and high-carrier-frequency RF processing.

The computing bottleneck in the baseband domain originates from the conflict between the low-latency demand of AI inference and the energy efficiency of electronic processors. Although electronic computing chips such as GPUs can provide theoretical peak throughput at the hundred-TOPS level, their usable computing power under the power constraints of a base station is far below the peak, making it difficult to guarantee computing latency [2]. Photonic computing accelerators address this challenge with a fundamentally different computing paradigm: the physical propagation of light in a medium inherently implements operators such as matrix multiplication and convolution, the operation latency is determined by the optical path rather than the clock frequency, and the energy efficiency can reach below 1 pJ/MAC—one to two orders of magnitude lower than that of GPUs[2], [3], [4].

The bandwidth bottleneck in the fronthaul domain is a structural problem inherent to the C-RAN architecture. CPRI transmits the complete I/Q sample stream in a transparent manner, rigidly binding bandwidth to antenna scale and sampling rate [5]; eCPRI, in turn, introduces higher system complexity and uncertain latency and jitter.

Toward 6G demands, the explosive growth of transmission bandwidth brought by multi-band, multi-channel, large-bandwidth communications places compound pressure on CPRI/eCPRI digital optical fronthaul[5]. Analog radio-over-fiber (A-RoF) modulates the RF waveform directly onto the optical carrier and transmits it in analog form, so fronthaul bandwidth is no longer rigidly coupled to sampling rate and the number of channels, and neither remote the ADC/DAC nor the digital IF chip are required, fundamentally eliminating the dual bottleneck of bandwidth and power consumption [6], [7] while effectively guaranteeing transmission latency.

The physical bottleneck in the RF domain intensifies as the frequency band rises. As 6G extends into millimeter-wave and terahertz bands, the efficiency of electronic devices degrades, electronic phase shifters produce beam squint under wideband signals, and the gain and noise performance of electronic mixers deteriorates significantly at high frequencies[8]. Microwave photonics exploits photonic technology to generate, transmit, and process microwave signals; the operating frequency of photonic devices is determined by the optical carrier frequency (about 200 THz) rather than carrier transit time, so link bandwidth does not decrease as the RF frequency rises [9], [10]. The fully photonic coherent radar system demonstrated in [11]verifies the feasibility of complete integration of photonic RF generation, beamforming, and down-conversion within a single system, establishing system-level technical confidence for the photonization of the RF front-end.

The Omni-PBS proposed in this paper aims to fill this gap. The Omni-PBS is a progressive evolutionary architecture that introduces photonic technologies into three functional domains—baseband processing, fronthaul transmission, and the RF front-end—and obtains system-level gains through cross-domain co-design. Its core idea is that the introduction of photonics is not a one-time all-optical transformation, but a deployment on demand and independent evolution in each functional domain, while cross-domain co-design minimizes optoelectronic conversion overhead and maximizes end-to-end energy efficiency. Rather than pursuing physical all-opticalization, this architecture takes system effectiveness as the optimization objective, allowing electronic processing to be retained where appropriate, and thereby exhibits stronger engineering feasibility and standards compatibility.

The main contributions of this paper are as follows. We propose the concept and three-domain architecture model of the Omni-PBS, establishing a unified photonization framework from baseband to antenna. We propose four co-design principles—end-to-end optical-domain continuity, cross-domain co-design, optical computing resource scheduling, and joint optimization of functional splitting—and reveal the mechanism by which the three domains produce system-level gains beyond single-domain summation. Building on the above concept and scheme, we analyze engineering constraints such as optical computing precision, analog link linearization, and photonic RF integration density, and discuss open problems and future directions.

The remainder of this paper is organized as follows. Chapter 2 defines the concept of the Omni-PBS and establishes the three-domain architecture model. Chapter 3 analyzes the technical foundations and communication applications of the optical baseband computing domain. Chapter 4 discusses the evolution of the analog optical fronthaul domain from the digital optical transmission to A-RoF. Chapter 5 elaborates the four core functions of the microwave-photonic RF domain. Chapter 6 analyzes the system-level co-design mechanism. Chapter 7 discusses engineering constraints, scenario fit, and open problems.

## 2 Omni-Photonic Base Station: Concept Definition and Architecture

### 2.1 Concept of the Omni-Photonic Base Station

The "fiber-wireless convergence" (FWC) conceptual framework proposed by Kitayama et al. [6]was the first to systematically discuss the unified application value of photonic technology in base stations, covering the co-design of RoF fronthaul, photonic RF processing, and wireless access. Lim et al. [7] subsequently provided a comprehensive survey of RoF technology in 5G and its evolution, covering system architecture, key challenges, and standardization progress, offering a panoramic view from devices to systems.

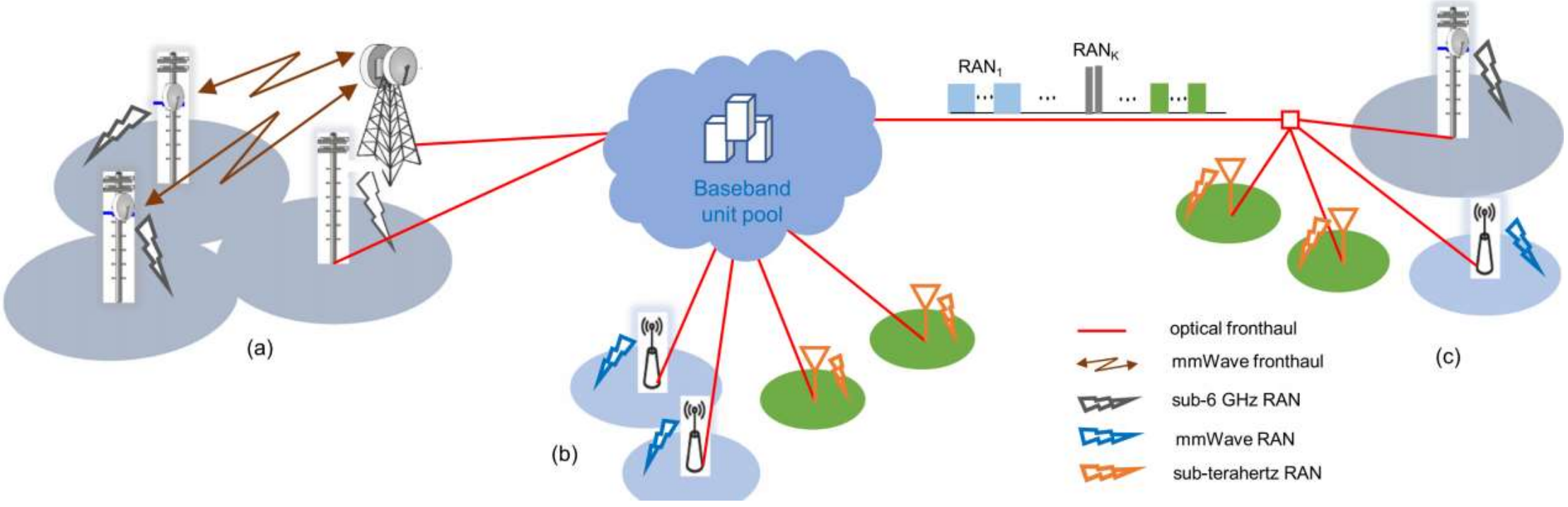

*Fig. 2.1 Fiber-wireless convergence system ([6] Dat et al.)*

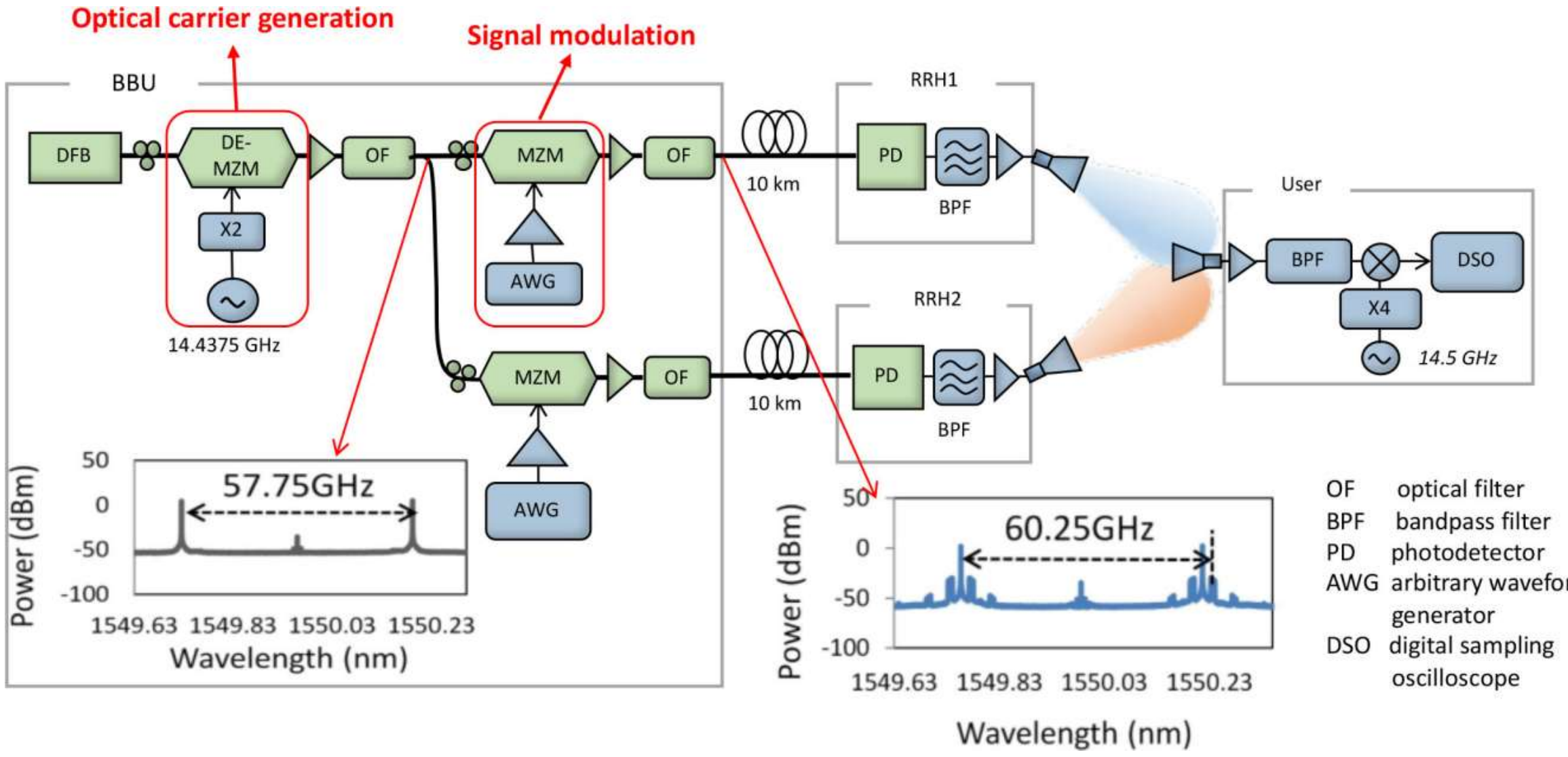

*Fig. 2.2 60 GHz RoF fronthaul experimental system ([7] Lim et al.)*

Building on the above work, this paper proposes for the first time, from the perspective of operators, the Omni-Photonic Base Station (Omni-PBS)—a progressive three-domain evolutionary architecture with optoelectronic synergy that introduces photonic technology into three functional domains: baseband processing, fronthaul transmission, and the RF front-end. Its core idea is that the introduction of photonics is not a one-time all-optical transformation, but an on-demand, independently evolving deployment in each functional domain, while cross-domain co-design yields system-level gains beyond single-domain summation.

This definition contains three layers of meaning. First, "Omni" emphasizes the ubiquity of photonics in the base station—from the antenna radiating element to the baseband processing unit, photonic technology permeates multiple stages of the signal chain rather than serving merely as a transmission medium. Second, "evolution" emphasizes gradualism—the photonization of each domain can advance independently of the others, is compatible with existing electronic base station architectures, and supports electro-optical hybrid operation. Third, "coordination" emphasizes cross-domain design—the three domains are coupled in function and complementary in performance, and a unified architecture minimizes optoelectronic conversion overhead and maximizes end-to-end energy efficiency.

It must be emphasized that the Omni-PBS does not pursue physical all-opticalization, but takes system effectiveness as the optimization objective, allowing electronic processing to be retained where appropriate. This pragmatic positioning endows the Omni-PBS with stronger engineering feasibility and standards compatibility.

## 2.2 Three-Domain Architecture Model of the Omni-PBS

The Omni-PBS partitions the traditional base station signal chain into three photonization-introducing domains: the optical baseband computing domain (Domain 1), the analog optical fronthaul domain (Domain 2), and the microwave-photonic RF domain (Domain 3). Physically, the three domains correspond to the Baseband Unit (BBU), fronthaul link, and RRU/antenna unit of the base station, and functionally they undertake computing, transmission, and RF processing, respectively. Fig. 2.3 shows the overall architecture of the Omni-PBS and the relationships among the three domains.

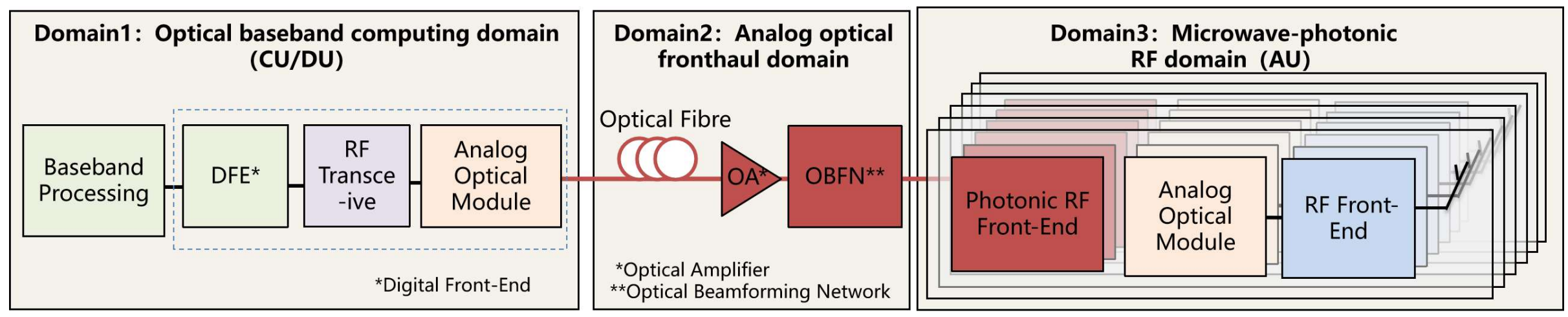

*Fig. 2.3 Overall architecture of the Omni-PBS*

Domain 1: Optical baseband computing domain. This domain resides on the BBU side and employs photonic accelerators to perform the linear computation-intensive tasks in baseband signal processing. The AI/ML air interface of 6G introduces a large amount of neural-network inference demand[1], in which matrix multiplication and convolution dominate the computation. Photonic accelerators—including Mach-Zehnder interferometer (MZI) arrays [3], microring resonator arrays, and diffractive optical networks ($D^2NN$) [12]—realize computation directly through the propagation and interference of light, with latency at the sub-nanosecond (hundreds of picoseconds) scale and no electrical energy consumed by the optical operation itself.

In the Omni-PBS architecture, the photonic accelerator and the GPU/CPU form a heterogeneous computing platform: linear operators (Conv2d, matrix multiplication, channel-matrix inversion) are offloaded to the photonic accelerator, while nonlinear operators (ReLU activation, normalization, nonlinear detection) remain in the electronic domain. The core functions of this domain include CSI compression and prediction based on neural networks [13], matrix operations in massive-MIMO detection, filtering operations in channel equalization[14], and beam prediction in beam management. The common characteristic of these tasks is a large computational load but a relatively tolerable precision requirement (6–8 bit equivalent precision), which precisely matches the precision and energy-efficiency profile of photonic accelerators [2].

Domain 2: Analog optical fronthaul domain. This domain connects the BBU to the remote unit and evolves the traditional digital optical fronthaul (CPRI/eCPRI [5], [15]) into analog optical transmission (A-RoF). A-RoF modulates the RF signal directly onto the optical carrier for transmission, moving the ADC/DAC and digital processing circuits from the remote unit to the central office; the remote unit then requires only optoelectronic conversion and RF amplification [16], greatly simplifying the complexity, power consumption, and cost of the remote unit.

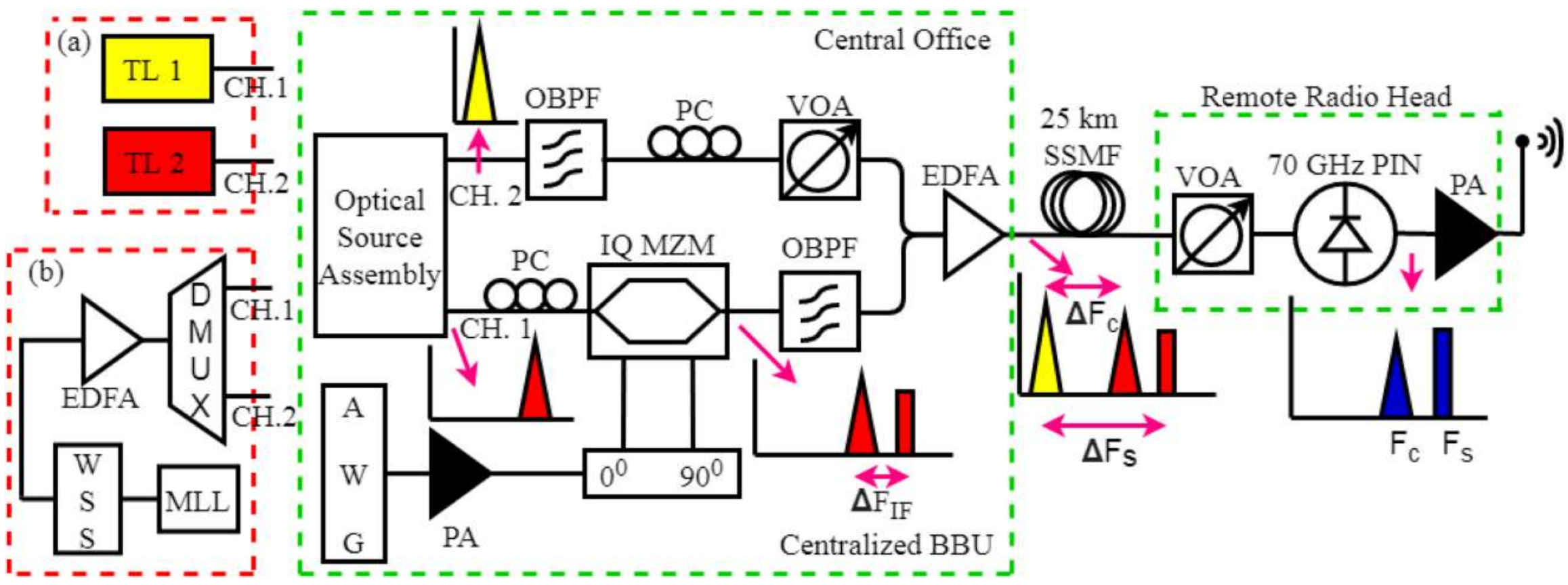

*Fig. 2.4 Analog heterodyne millimeter-wave A-RoF fronthaul link ( [17] Delmade et al.)*

This domain must balance transmission performance against deployment flexibility: A-RoF faces nonlinear distortion (laser RIN, modulator IM2/IM3, dispersive power fading) and dynamic-range limitations [18], which must be compensated by digital predistortion and optical-domain equalization.

Regarding functional splitting, A-RoF shows more pronounced advantages under low-layer splitting (Split 7.2x and below), because the transmitted signal is then closer to the analog RF domain; under high-layer splitting the remote unit retains digital processing, and the digital optical transmission remains applicable. Therefore, Domain 2 adopts a hybrid the digital optical transmission/A-RoF mode in the early evolution stage and gradually transitions to A-RoF dominance.

Domain 3: Microwave-photonic RF domain. This domain resides in the remote unit and uses microwave-photonics technology to realize the generation, transmission, processing, and control of RF signals. The RF front-end in this domain typically incorporates filter banks, switches, and other RF signal conditioning modules. Compared with traditional electronic RF front-ends, the photonic RF front-end has irreplaceable advantages in three key functions: high-precision ultra-wideband analog beamforming, filtering and interference management, and frequency generation and conversion.

In beamforming, the optical true-time-delay (TTD) array fundamentally eliminates the beam squint caused by electronic phase shifters under wideband signals[11]. As 6G extends into millimeter-wave and terahertz bands, the instantaneous signal bandwidth can reach several GHz, and the frequency-dependent phase error of electronic phase shifters becomes a bottleneck; TTD realizes frequency-independent delay through physical optical-path differences, making it one feasible scheme for high-precision ultra-wideband beamforming.

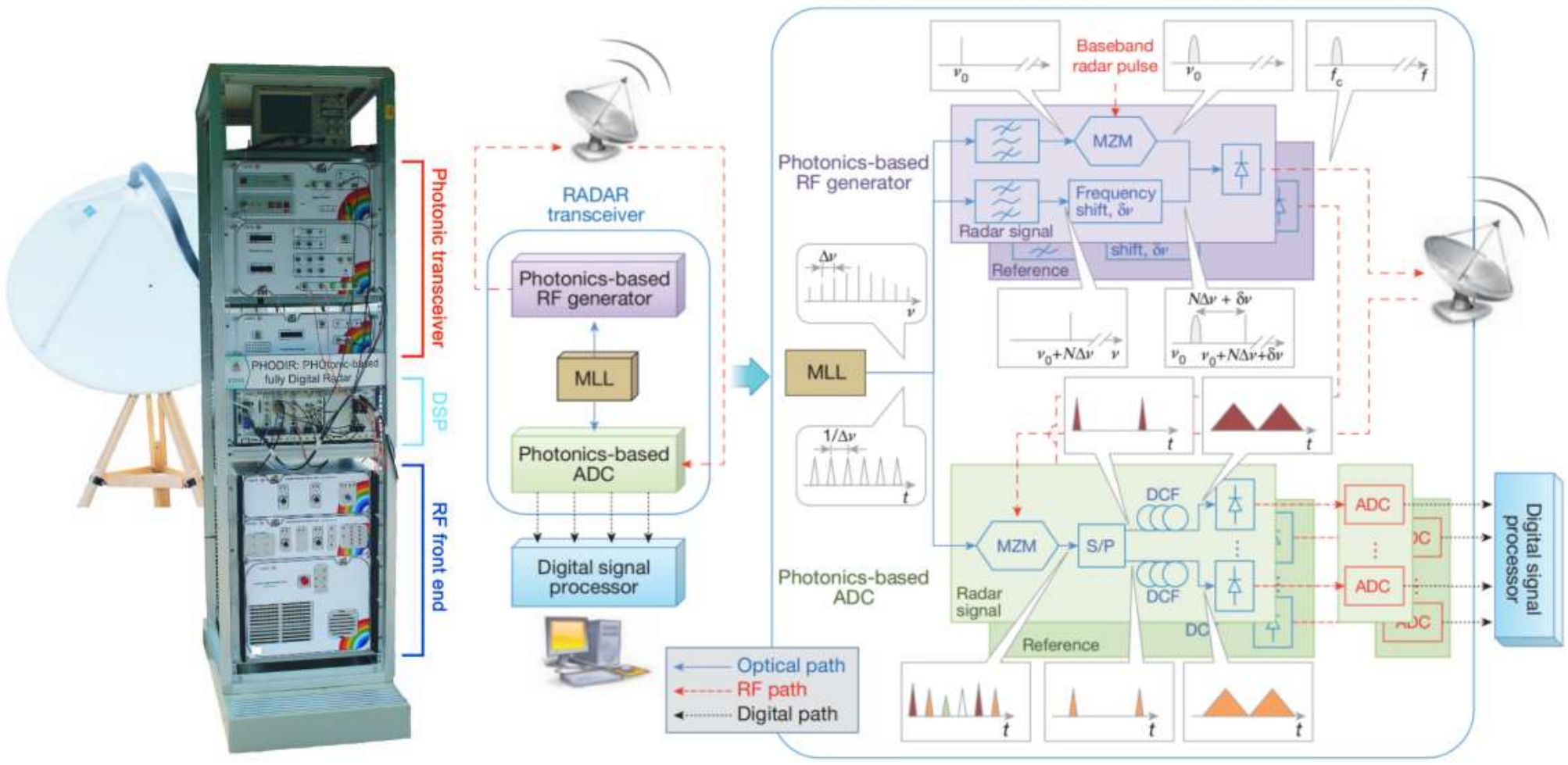

*Fig. 2.5 Fully photonic coherent radar system (PHODIR) ( [11] Ghelfi et al.)*

In filtering and interference management, programmable microwave-photonic filters perform RF selective filtering in the optical domain, with nanosecond-level reconfiguration speed and GHz-level tuning range [19], and can dynamically suppress co-channel and adjacent-channel interference—a capability especially important for 6G integrated sensing and communication (ISAC) scenarios.

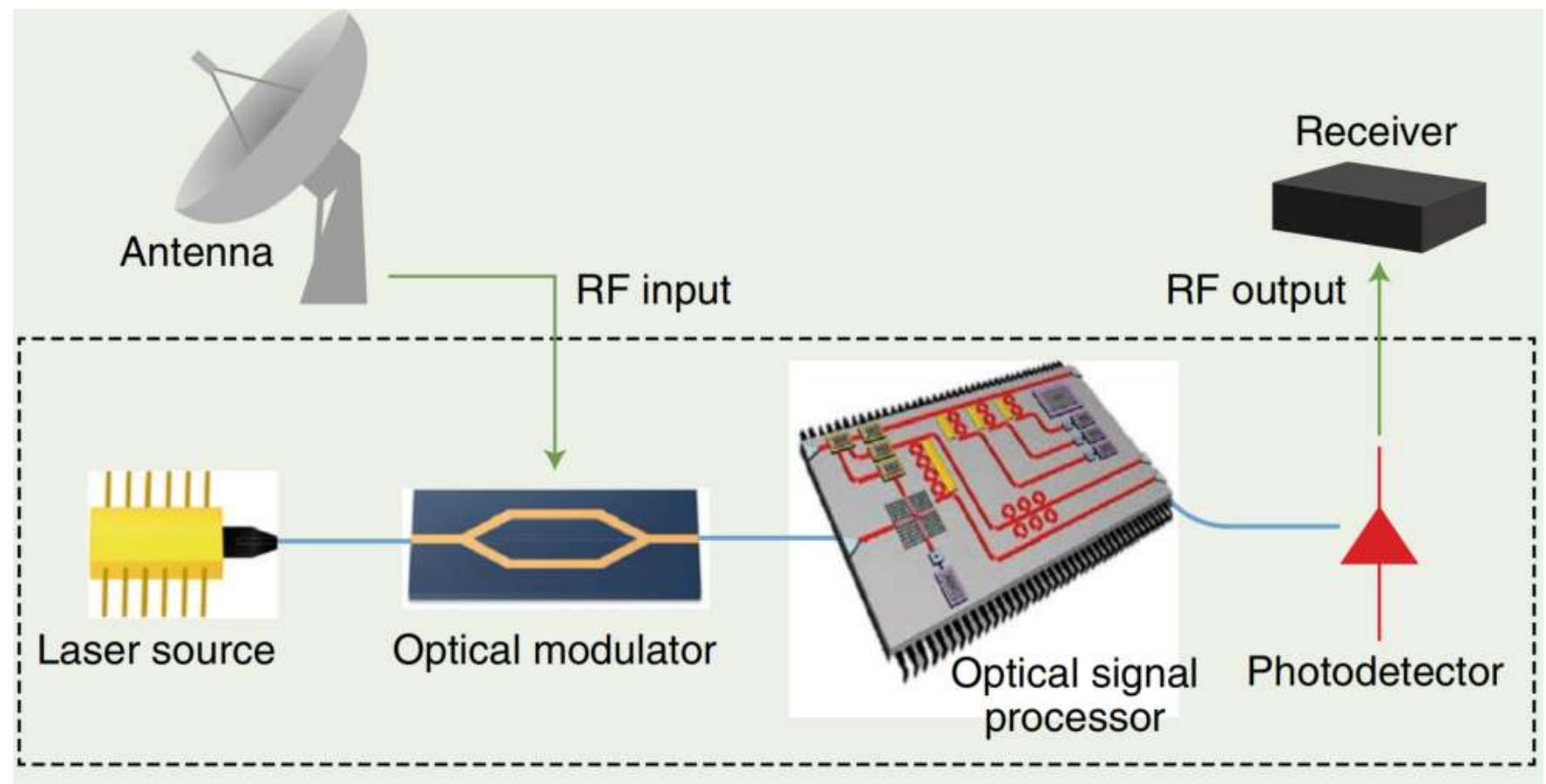

*Fig. 2.6 Integrated microwave-photonics system architecture ( [19]Marpaung et al.)*

In frequency generation and conversion, optical heterodyne and optical frequency combs can generate millimeter-wave/terahertz carriers with low phase noise [20], and photonic mixers provide a wideband down-conversion alternative [21]. Photonic local-oscillator (LO) distribution technology can allocate the same high-purity optical carrier to multiple remote units, realizing distributed coherent coordination[11].

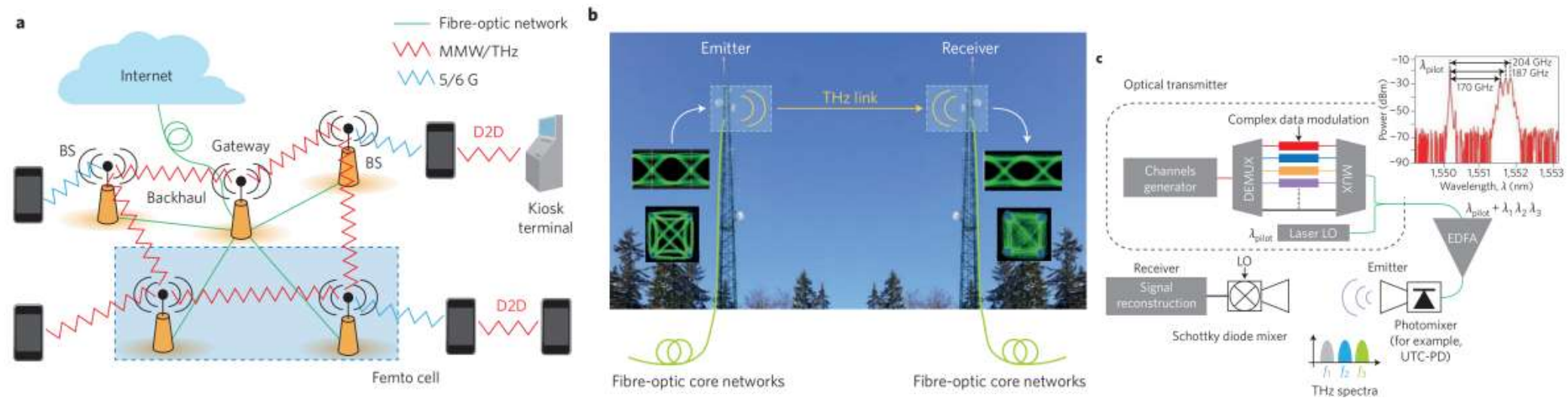

*Fig. 2.7 THz network applications and photonic technical solutions ([20] Nagatsuma et al.)*

The fusion of Domain 3 and Domain 2 is a key design point of the Omni-PBS architecture: when A-RoF delivers the analog RF signal directly to the remote end, the photonic RF front-end performs beamforming and down-conversion in the optical domain, and then only a single optoelectronic conversion is needed to drive the antenna radiation. This "optical-domain continuity" path eliminates the O/E–E/O conversion between the fronthaul and the RF front-end in traditional architectures, and is the core value of cross-domain coordination, to be analyzed in detail in Section 2.5.

# 3 Implementation of the Optical Baseband Computing Functional Domain

## 3.1 AI Computing Demands of the 6G Baseband and Bottlenecks in Existing Base Station Architectures

6G elevates AI/ML from an optional enhancement to a native air-interface capability. In TR 38.812, 3GPP Rel-19 defines three categories of AI/ML use cases: CSI compression and feedback, beam management (beam prediction/beam recovery), and positioning enhancement [1]. Rel-20 further extends AI/ML to channel coding, hybrid automatic repeat request (HARQ), and scheduling optimization. The common characteristics of these use cases are heavy online inference load, strict latency constraints (physical-layer AI inference must complete within microseconds), and models dominated by fully connected and convolutional networks [2].

However, existing base station architectures face fundamental bottlenecks when hosting these AI inference workloads. Current baseband units (BBUs) primarily rely on GPUs and dedicated ASICs for AI acceleration, but GPUs suffer from a severe power wall — a single high-end GPU already exceeds 400 W, while the power budget of a single base station cabinet in dense deployment scenarios is highly constrained. Moreover, the pipeline latency introduced by GPU batch-processing optimization strategies stands in fundamental conflict with the low-latency requirements of real-time physical-layer inference: increasing batch size to improve throughput introduces millisecond-scale queuing delays, whereas physical-layer AI inference demands microsecond-scale response. In addition, although ASICs offer high energy efficiency, their fixed logic design cannot accommodate the rapid iteration of AI models in 6G — changes in channel conditions, standards evolution, or scenario migration all require model updates, while the hardware rigidity of ASICs leaves them without the necessary flexibility. This "power-latency-flexibility" trilemma severely constrains both the deployment density and the performance ceiling of AI capabilities on the base station side.

Photonic computing accelerators can address this challenge with a fundamentally different computing paradigm: the physical propagation of light in a medium inherently implements matrix multiplication, the operation latency is determined by the optical path rather than the clock frequency, and a single 64×64 matrix multiplication takes only about 5 ns [22]. More importantly, the propagation of light consumes almost no energy; the energy is mainly concentrated in electro-optic modulation at the input and photoelectric detection at the output [2], yielding an energy efficiency below 1 pJ/MAC—one to two orders of magnitude lower than that of GPUs.

### 3.2 Hardware Implementation of Photonic Computing and Its Applications in Communications

The physical basis of photonic computing lies in the fact that the propagation of light in a medium naturally corresponds to linear transformation operations. In 2017, the programmable nanophotonic circuit proposed by Shen et al. [22] first demonstrated that optical matrix multiplication through cascaded MZI interferometer arrays can directly execute machine-learning inference, completing a vowel-classification task at the speed of light.

In communication-specific applications, preliminary explorations exist. Photonic reservoir computing exploits the nonlinear dynamics of photonic devices to process time-series signals, showing advantages in channel equalization. Zhang et al. [23] applied photonic reservoir computing to wireless channel equalization, realizing real-time compensation for multipath channels at a rate of 100 Gbaud, with a processing speed far exceeding that of electronic DSP schemes.

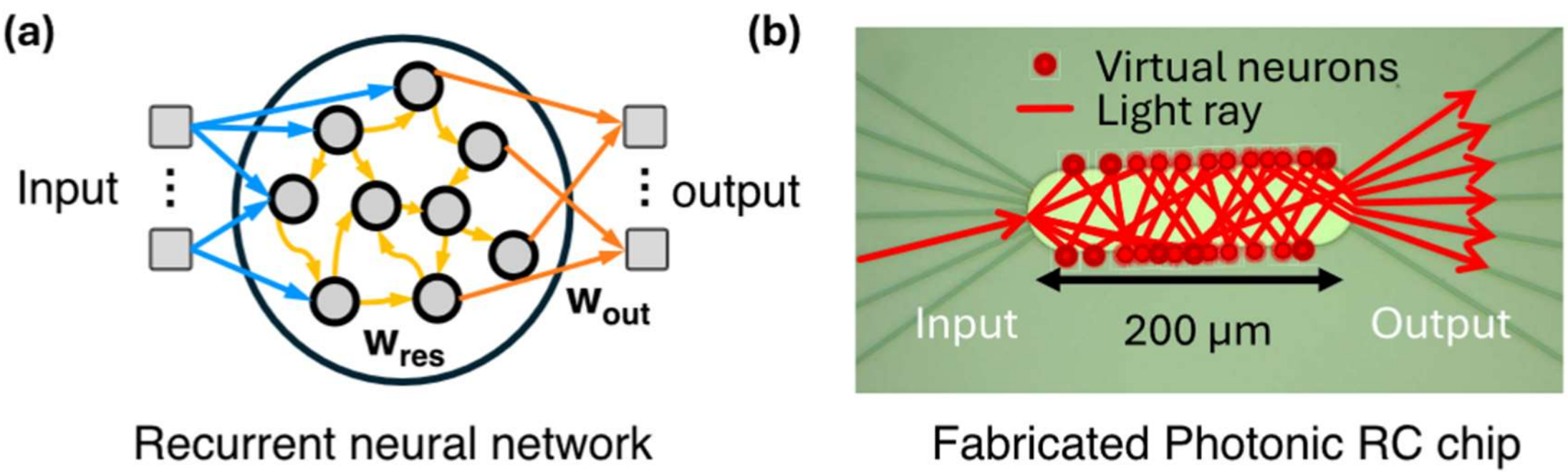

*Fig. 3.1 Structure and chip photograph of the reservoir-computing-based recurrent neural network*

Yu, Xu, and Zou et al. [24] proposed a low-complexity photonic wireless channel-estimation architecture based on matrix condition-number theory, reducing the hardware complexity of OFDM systems by tens to hundreds of times, and experimentally verified channel state information mean-square error at the $10^{-4}$ level.

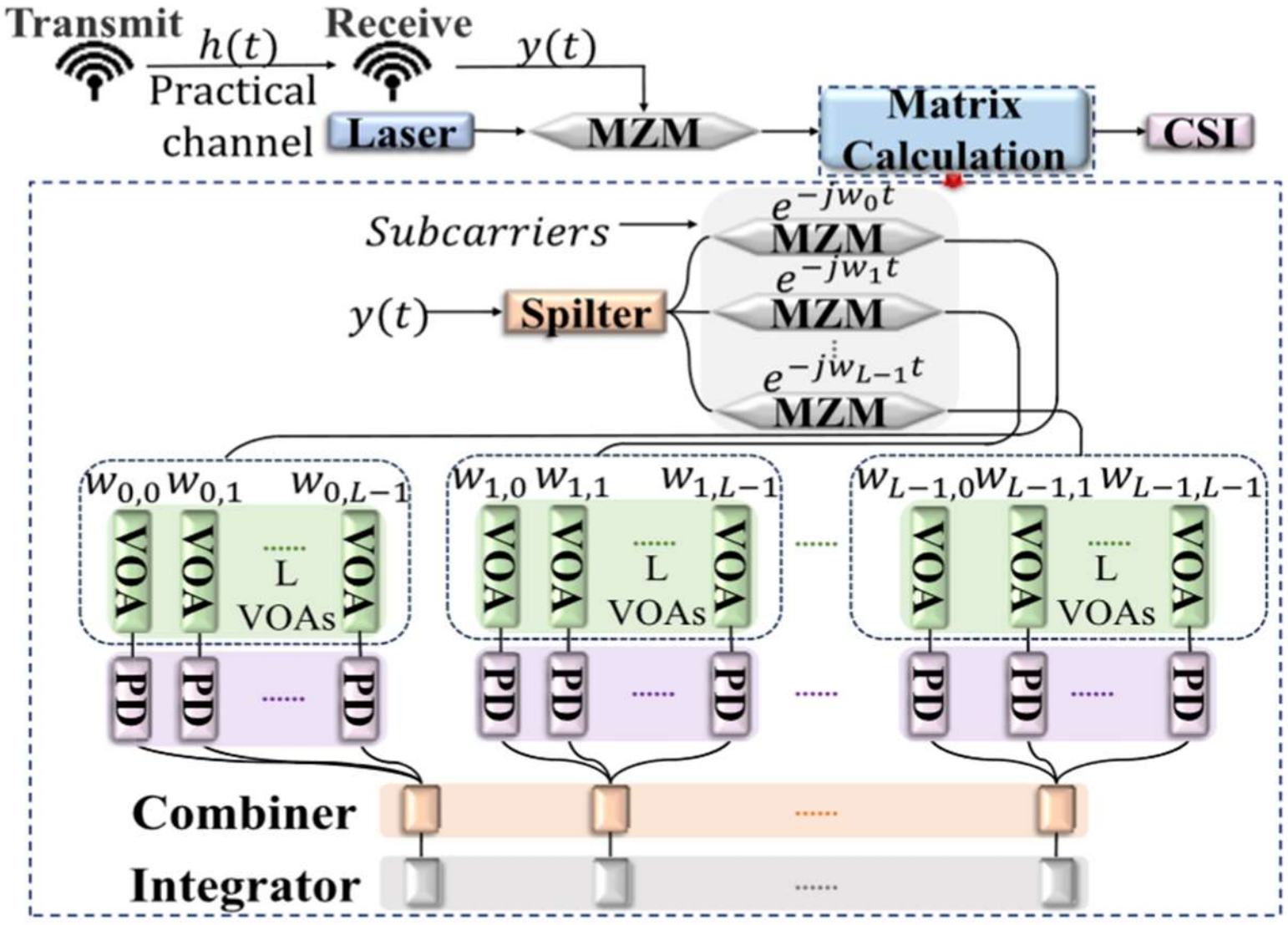

*Fig. 3.2 Conceptual architecture of the simplified photonic channel-estimation method*

In the latency dimension, the fully integrated hybrid photonic processor implemented by Khaled et al. [25] achieves a processing delay of only 30 ps, verified in real-time descrambling of 5 Gb/s MIMO signals, and its picosecond-level response capability is well suited to real-time MIMO and RF applications. These works indicate the application potential of optical computing in communication baseband processing, but existing research mainly targets single-function verification, and end-to-end integration with the baseband processing flow of base stations has not yet been carried out.

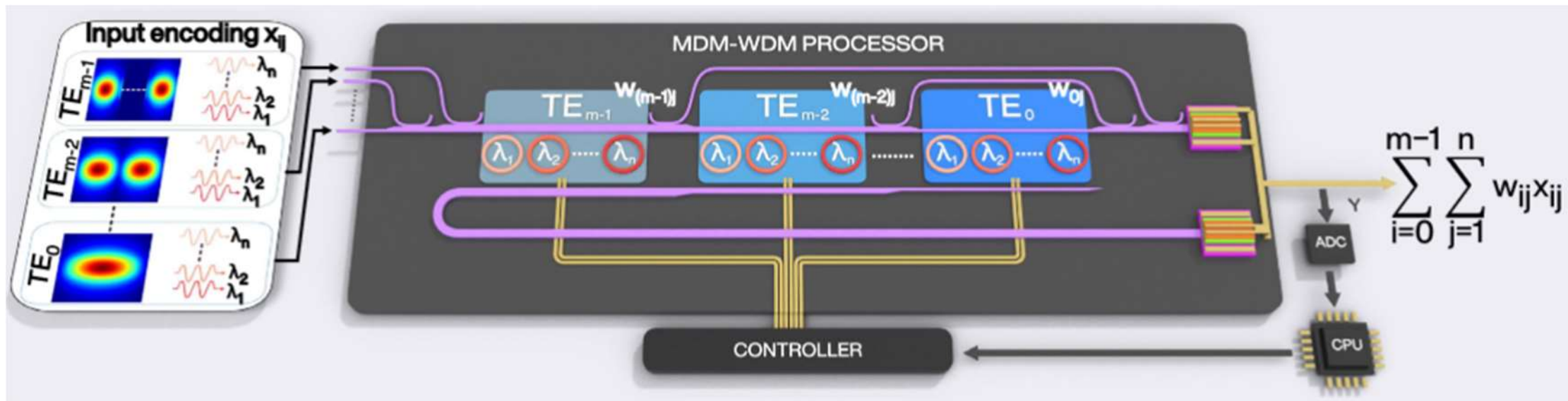

*Fig. 3.3 Schematic of a large-scale fully integrated hybrid multi-mode multi-wavelength photonic processor*

### 3.3 Photonic Computing Hardware and Operator-Mapping Scheme

Building on the premise that photonic computing hardware can effectively support various computing tasks, the implementation idea of the Omni-PBS is to map the various linear operators in the baseband AI inference flow onto photonic computing accelerators for accelerated operation. The mainstream photonic computing accelerators differ in their optical computing principles and hardware characteristics, leading to significantly different mapping schemes for baseband linear operators.

The MZI interferometer array implements general matrix–vector multiplication through a programmable optical scattering matrix. The PACE chip from Lightelligence adopts a 64×64 MZI mesh, supporting INT4 weight precision and INT8 output precision[26]. Its computation proceeds as follows: the input vector is encoded into optical intensity through electro-optic modulation, performs the linear transformation corresponding to the unitary-matrix decomposition through the optical interference of the MZI mesh, and the result is read out through a photodetector array. The single-operation latency is about 5 ns, with an equivalent throughput of about 8.19 TOPS.

The diffractive deep neural network ($D^2NN$) uses on-chip diffraction to implement fully connected layer computation. The Taiji architecture of Tsinghua University implements million-neuron-scale fully connected inference through cascaded diffractive layers, where the phase modulation of each diffractive layer corresponds to the multiplication of the weight matrix [27]. The advantage of D²NN lies in large-scale parallelism and zero power consumption of passive computing, but the number of layers is limited by noise accumulation, and the weights are fixed and difficult to reconfigure at inference time.

The microring resonator array uses wavelength-division multiplexing to implement convolution. By multiplying each wavelength component of the input signal with the microring weight and summing, one-dimensional convolution is realized. Feldmann et al. [4] achieved a convolution rate at the T-MAC/s level on an integrated photonic chip, an architecture directly applicable to one-dimensional filtering tasks such as channel equalization.

The parallel optical convolution streaming processor (OCSP) based on a microcomb uses wavelength-division multiplexing and time–space interleaving, and implements multi-tap finite impulse response (FIR) filtering through cascaded Mach-Zehnder interferometers (MZIs), performing parallel convolution on data streams of multiple wavelength channels simultaneously. Wang et al. [28] implemented an 8-tap, 50 GBaud convolution chip on a standard SOI platform, with a peak computation speed of 4 TOPS (5-wavelength parallel) and an on-chip self-calibration mechanism to precisely regulate the convolution weights. Table 1 compares the key parameters of the above photonic computing accelerators and their applicable scenarios in baseband processing.

*Table 1: Comparison of photonic computing accelerator architectures and baseband application mapping*

| **Parameter** | **MZI array [22]** | **D²NN [27]** | **Microring convolution [4]** | **MZI convolution [28]** |
|---|---|---|---|---|
| Computing operator | Matrix–vector multiplication | Fully connected layer | 1-D convolution | 2-D convolution |
| Matrix scale | 64×64 | ~$10^6$ neurons | ~40 channels | ~5 channels |
| Precision (ENOB) | 6–8 bit | 4–6 bit | ~8 bit | — |
| Reconfigurability | Reconfigurable | Fixed weights | Reconfigurable | Reconfigurable |
| Latency | ~5 ns | ~10 ps | ~100 ps | ~290 ps |
| Applicable baseband task | CSI compression, MIMO detection | Beam prediction (fixed model) | Channel equalization, filtering | Channel equalization, inter-symbol interference compensation |

Because analog photonic computing is constrained by device non-idealities, there is an inherent precision limitation that must be accommodated to the precision requirements of baseband inference tasks through precision quantization and noise-simulation compensation. In the model training and deployment stages, the original FP32 high-precision weights and activations must be quantized and adapted to the effective precision range of the photonic computing hardware; among these, the MZI array adapts an INT4-weight, INT8-output fixed-point quantization scheme [26], maintaining good inference performance at low bit-width operation; the analog effective precision of the D²NN diffractive optical network is only 4–6 bit, requiring quantization-aware training to counteract quantization error and ensure controllable model inference precision [2]. At the same time, photonic computing systems suffer inherent device noise such as MZI phase drift, photodetector shot noise, and modulator nonlinearity; to accurately evaluate actual deployment performance, the compiler can introduce an equivalent noise model in the simulation stage to quantitatively analyze the precision-loss pattern of photonic computing inference.

Taking the general residual network ResNet18 as an example, thanks to the strong robustness of its residual connections and ReLU activations against quantization error, INT8 quantization reduces inference latency by 25%–30% with less than 1% precision loss: on the GPU (TensorRT) platform the inference latency drops from 0.95 s to 0.71 s (a 25.3% reduction, with only 0.7% precision loss); on the CPU (TensorFlow Lite) platform it drops from 27.8 s to 19.2 s (a 30.9% reduction, with only 0.1% precision loss), providing support for the precision-controllable deployment of baseband inference tasks on photonic computing hardware.

Accordingly, this paper constructs a simulation model for wireless AI inference: based on the classic ResNet18, constrained by the optical convolution hardware (the convolution kernel size does not exceed the square root of the number of light-source wavelengths; this work uses a 9-wavelength light source, corresponding to a 3×3 convolution kernel), the first 7×7 convolution is replaced by a 3×3 convolution while the remaining convolution kernels are unchanged, so that each convolution layer naturally fits the optical convolution unit. The validation task is wireless AI inference based on interference frequency-domain image recognition, using more than 9,000 interference records from the communication OMC northbound management platform, comparing 1,000 data points (10 interference classes in total) along four dimensions: inference latency, precision, resource overhead, and quantization effect.

*Table 2: Simulation results of INT8 quantization based on ResNet18 (expected)*

| Quantization method | Platform | Inference latency | Latency reduction | Accuracy |
|---|---|---|---|---|
| FP32 (baseline) | GPU (TensorRT) | 0.95 s | — | 95.3% |
| INT8 | GPU (TensorRT) | 0.71 s | −25.3% | 94.6% |
| FP32 (baseline) | CPU (TensorFlow Lite) | 27.8 s | — | 95.3% |
| INT8 | CPU (TensorFlow Lite) | 19.2 s | −30.9% | 95.2% |

The simulation results show that INT8 quantization reduces inference latency by 25.3% on the GPU platform and 30.9% on the CPU platform with less than 1% precision loss, verifying the feasibility of maintaining inference performance under the precision constraints of photonic computing; combined with the earlier electro-optical comparison showing that optical convolution reduces convolution-layer latency by about 90.86% and overall inference latency by about 85.51%, with power consumption dropping from 250 W to 1 W, the quantization compensation mechanism and equivalent-noise simulation together constitute the empirical basis for precision-controllable and energy-efficiency-controllable deployment of baseband inference tasks on photonic computing hardware.

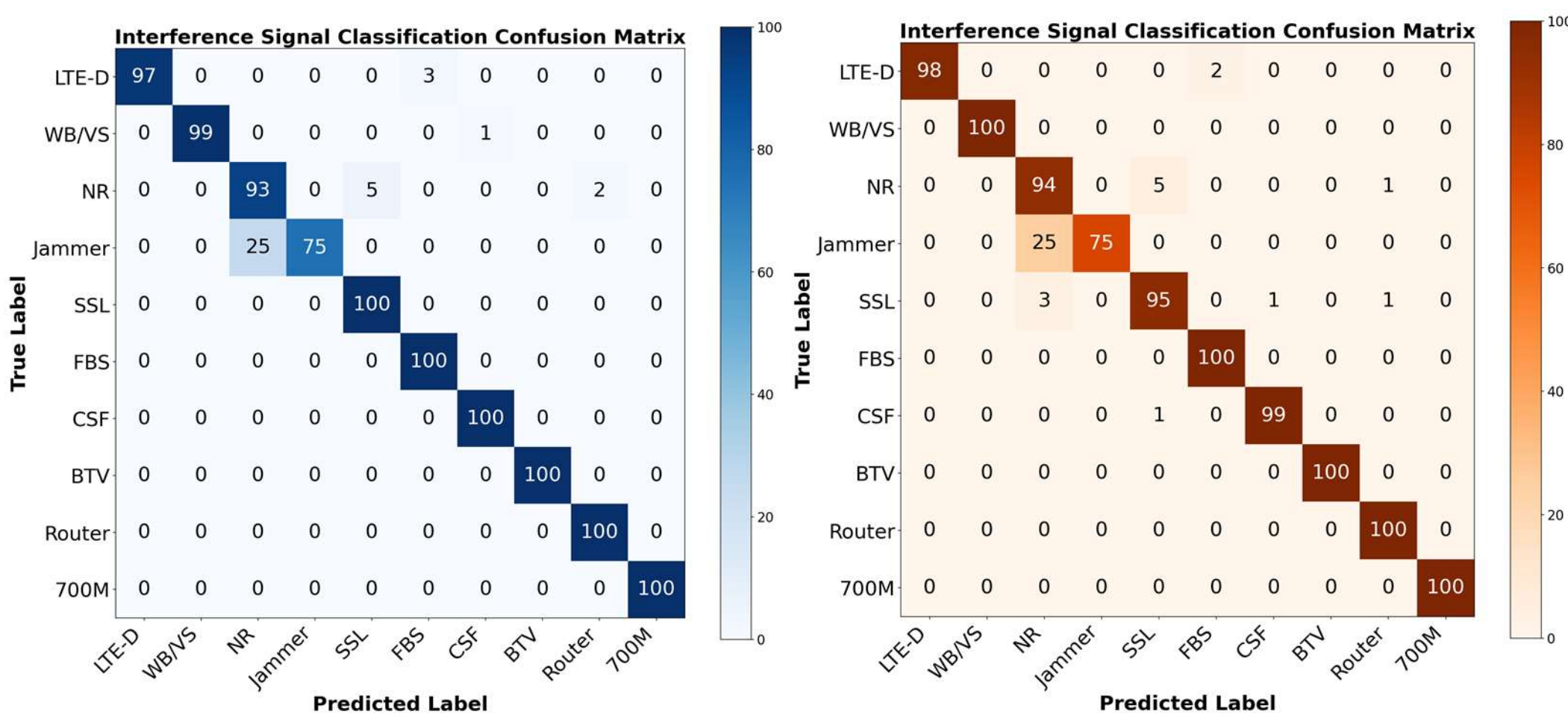

*Figure 3.4 Confusion matrices corresponding to simulation results of INT4 (left) and full precision (right)*

# 4 Analog Optical Fronthaul Domain

## 4.1 Bottlenecks Faced by Digital Fronthaul in the 6G Era

The evolution of base station architecture—from distributed to centralized, from closed to open—has always revolved around one core contradiction: the trade-off between centralization gain and fronthaul cost. The traditional 4G base station adopted an architecture that separates the baseband processing unit (BBU) from the remote radio unit (RRU), connected via the CPRI interface [5]. The BBU hosts all baseband processing (L1/L2/L3), and the RRU is responsible for RF transceiver and fronthaul interface adaptation; CPRI transmits the complete I/Q sample stream in a "dumb-pipe" manner. CPRI bandwidth is rigidly bound to antenna scale and sampling rate, and the interface between BBU and RRU is vendor-proprietary, making multi-vendor interoperability difficult.

The Cloud-RAN (C-RAN) architecture centralizes baseband processing [29]. Multiple BBUs converge into a baseband pool, enabling dynamic sharing of computing resources and multi-point coordination (CoMP); RRUs are deployed in a distributed manner to cover the service area. The core value of C-RAN lies in baseband resource sharing and coordination gain, but its fronthaul bandwidth demand is extremely rigid due to CPRI's transparent transmission nature. After 5G introduced massive MIMO, a 64×64 array at 100 MHz bandwidth requires over 100 Gbps of CPRI bandwidth [5], consuming enormous fiber resources.

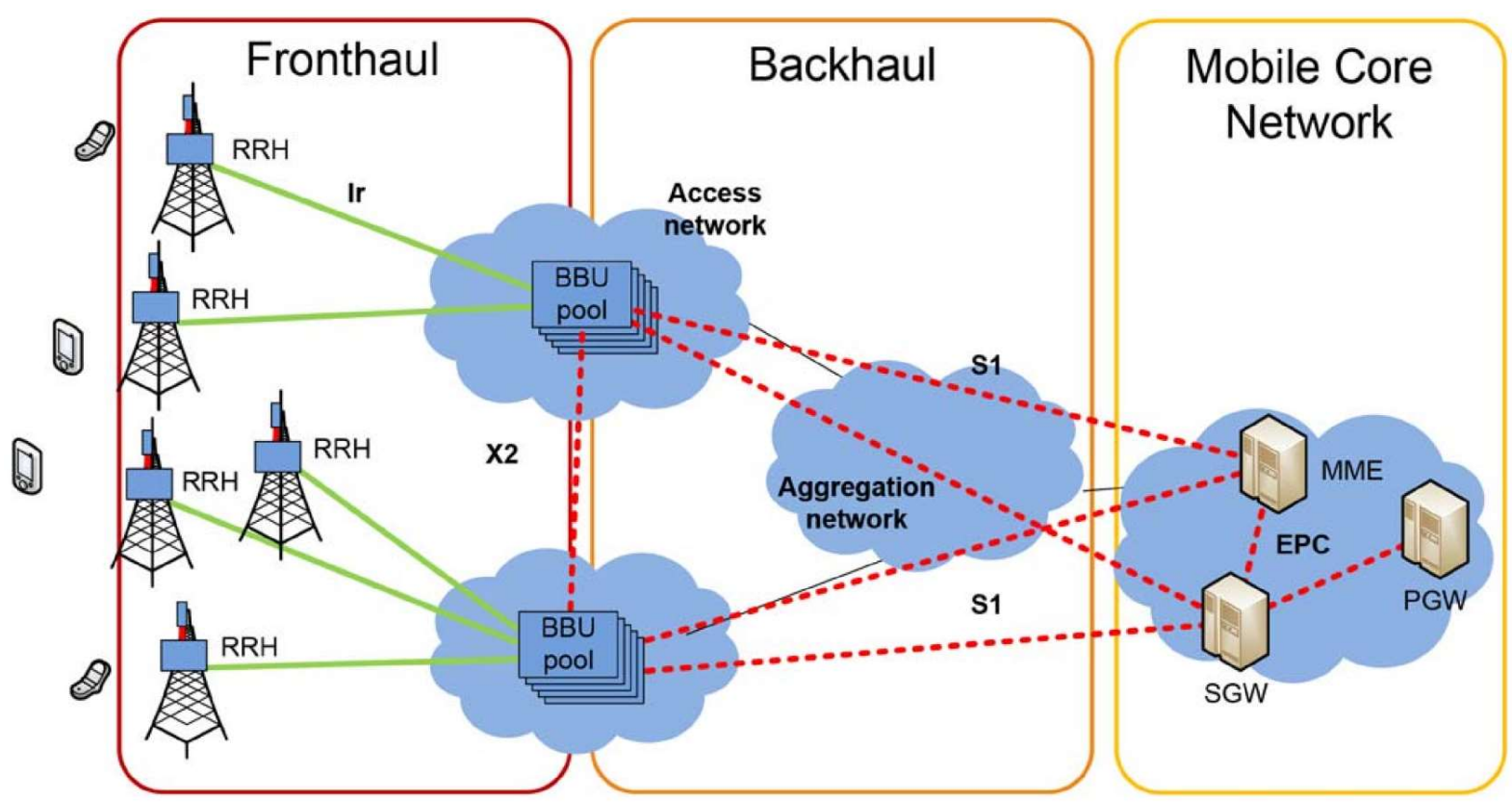

*Fig. 4.1 C-RAN architecture schematic ( [29] Checko et al.)*

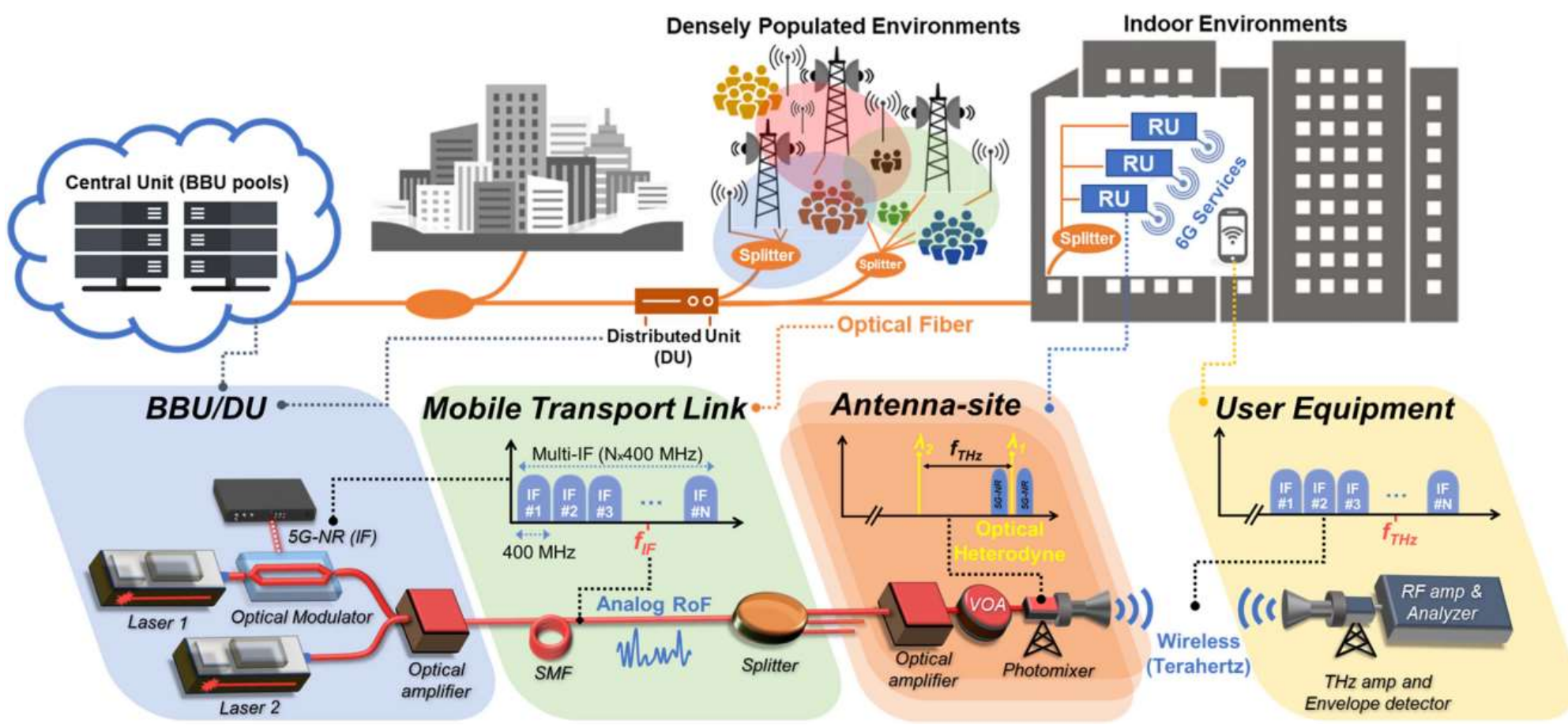

*Fig. 4.2 Photonic THz communication system architecture based on RoF (after [30] Sung et al.)*

However, the above architectural evolution has always been based on one invariant premise: the fronthaul link operates in the digital domain. The digital optical transmission paradigm—digital I/Q transmission after ADC/DAC sampling—rigidly couples fronthaul bandwidth to antenna scale and sampling rate. Toward 6G, antenna arrays are expected to expand from 64 channels to 256 or even 512 channels, with instantaneous bandwidth at the GHz level, and the digital optical transmission fronthaul faces a compound bandwidth explosion; meanwhile, the power consumption of high-speed ADC/DAC under large-scale arrays is approaching engineering limits [31].

### 4.2 A-RoF Technology Implementation

The evolution from the digital optical transmission to A-RoF is not achieved overnight; it must consider the compatibility of existing deployments and operational continuity. The Omni-PBS adopts a three-stage transition strategy.

The first stage is the hybrid mode: the digital optical transmission and A-RoF coexist in the same fronthaul network. Newly built RRUs adopt the A-RoF interface to verify technical feasibility and operational experience, while legacy RRUs retain the digital optical transmission; the BBU side is equipped with dual-mode fronthaul interface cards, enabling flexible switching between the two modes through software definition.

The second stage is A-RoF dominance: newly built RRUs fully adopt A-RoF, and the digital optical transmission is retained only in specific scenarios unsuitable for analog transmission, such as long distances (>20 km) or high-reliability requirements. The deployment data accumulated in this stage provides practical support for subsequent standardization.

The third stage is A-RoF standardization:A-RoF interface parameters (including modulation format, wavelength planning, monitoring channels, and management-plane protocols) are formed into a unified industry standard, enabling interoperability between different equipment vendors and completing the paradigm shift of fronthaul technology from "digital and proprietary" to "analog and open."

### 4.3 A-RoF Architecture Implementation Solution

Based on the above analysis of A-RoF technology, this section proposes an A-RoF implementation scheme for 6G at the architectural level. Its core idea is to centrally deploy all functions—baseband processing, DPD linearization, and RF generation—on the central-office side, while the remote end degrades into a minimal unit requiring only optoelectronic conversion and RF transmission.Since the multi-channel, large-bandwidth digital intermediate-frequency (IF) chips and transceiver chips are integrated as a resource pool in the BBU with a relatively favorable working environment, the power consumption and complexity of the RRUs operating in harsh outdoor environments are greatly reduced. In particular, for distributed pico base stations, the number of front-end pRRUs and RF channels is often far greater than the number of BBUs (including HUBs) and digital channels; therefore, the centralized pooling of baseband and intermediate-frequency circuits, combined with the significant simplification of remote units, brings greater cost and energy-consumption reductions to the entire base station.

At the BBU side, multiple RF channels are assigned independent wavelengths and modulated onto respective optical carriers, then injected into the fiber link through a wavelength-division multiplexing (WDM) combiner. At the RRU side, the optical carriers demultiplexed by WDM are directly injected into a uni-traveling-carrier photodiode (UTC-PD); benefited from its high saturated optical power (up to +20 dBm or above), the output RF power can directly drive the antenna without a power amplifier [32]. Vega et al. have demonstrated a compact Ka-band transmitter scheme with UTC-PD directly connected to the antenna, confirming the feasibility of this architecture in the millimeter-wave band [33].When the number of channels is large, wavelength-division multiplexing/demultiplexing combined with RF mixing and combining/splitting filters can be adopted to reduce the number of fibers required.

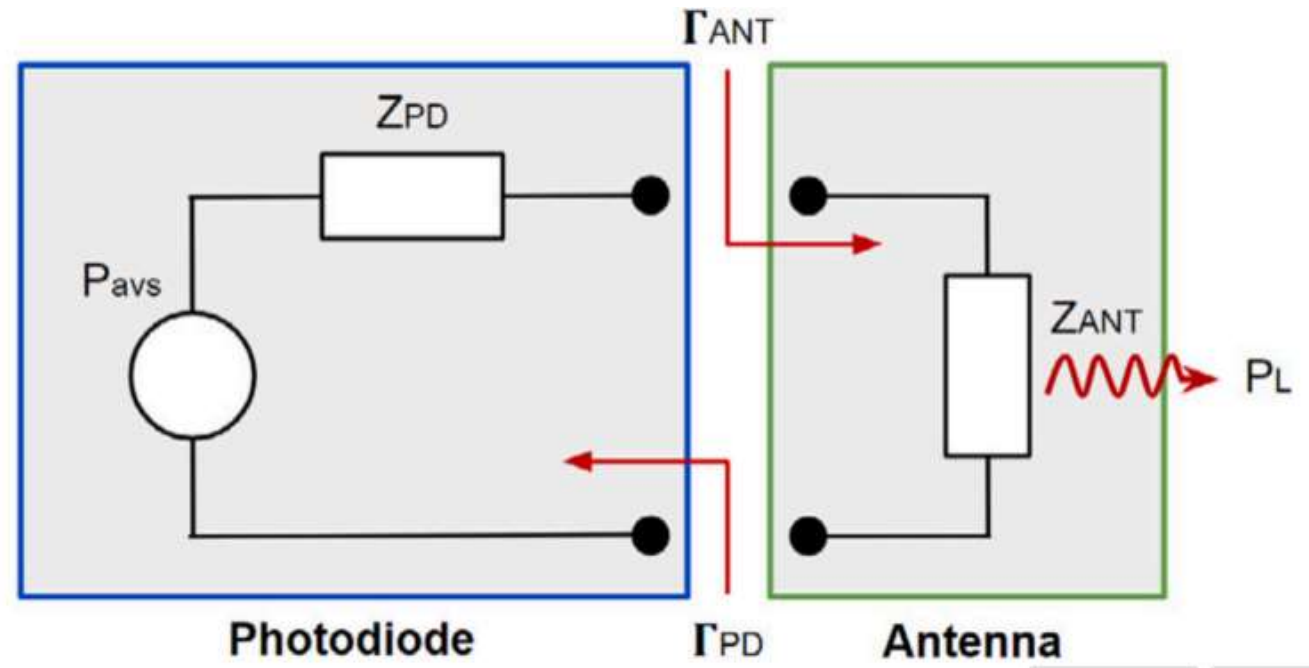

*Fig. 4.3 Equivalent circuit of UTC-PD directly connected to the antenna (*[33] *Vega et al.)*

The above architecture can be summarized as two changes: first, all digital processing is physically centralized at the BBU side, leveraging the ample power supply and heat dissipation of the equipment room to support high-complexity algorithms, and extending the DPD compensation target from the PA to the end-to-end optical link; second, a high-power UTC-PD realizes RF-direct-drive of the antenna at the RRU side, reducing the number of active devices at the remote end by 50%. Puerta et al.'s recent system-level evaluation of A-RoF shows that the architecture can meet 3GPP EVM and ACLR requirements in a 6G D-MIMO network and obtains a 9.4 dB MIMO gain in four-transmitter coordinated transmission [34].

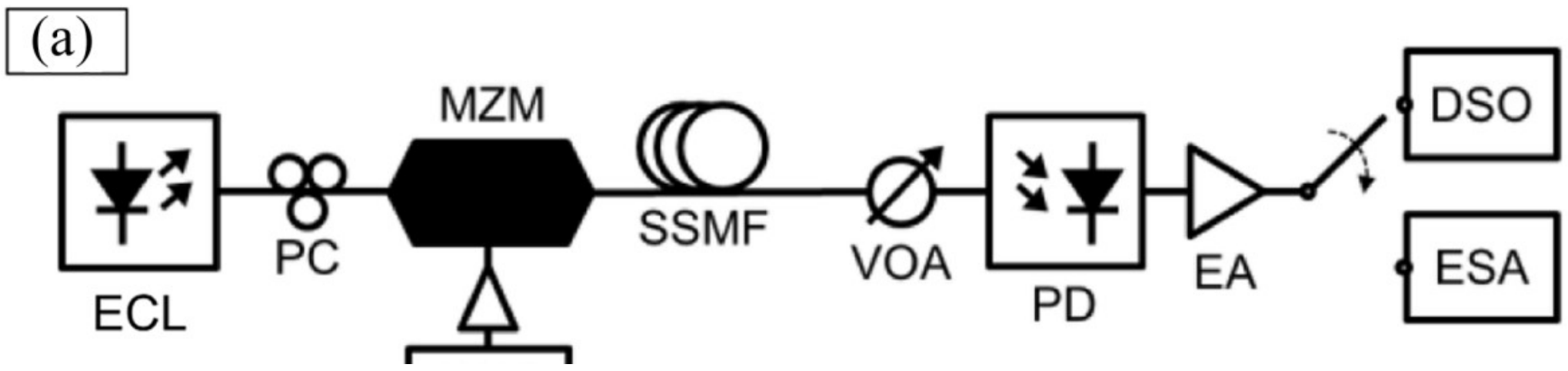

*Fig. 4.4 A-RoF experimental system block diagram (* [34] *Puerta et al.)*

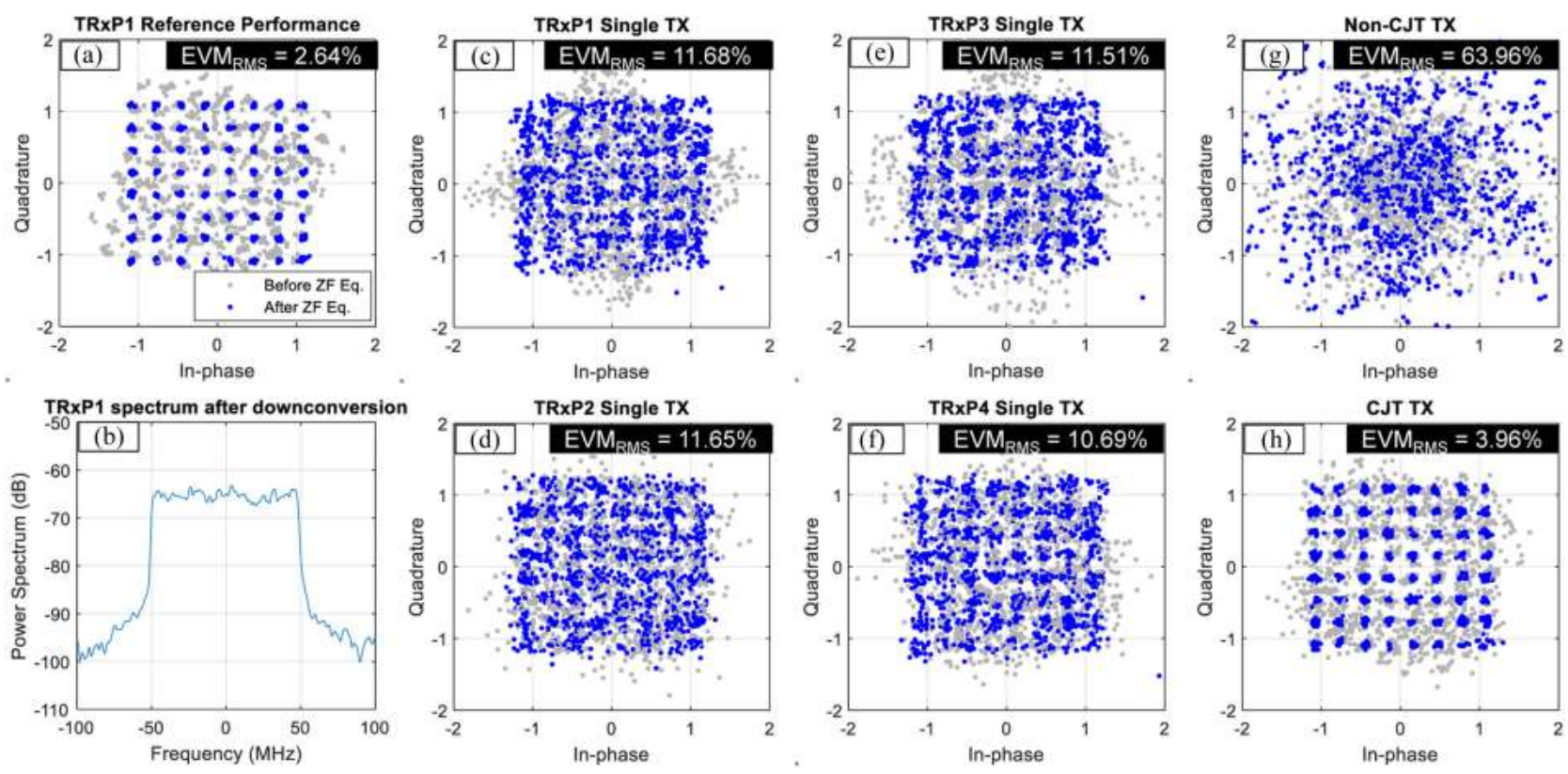

*Fig. 4.5 A-RoF four-transmitter MIMO constellation and spectrum ($EVM_{RMS}$ = 1.4%–2.6%) (* [34]*Puerta et al.)*

To validate the practical performance of the analog optical fronthaul link in the proposed A-RoF architecture, we built an experimental link based on a directly modulated laser (DML) and a UTC-PD. The test setup, shown in Fig. 4.8, consists of a signal generator, a DML, a fiber patch cord, a UTC-PD, a low-noise amplifier (LNA), and a spectrum analyzer. A single-tone signal was generated and the system noise floor was raised through the LNA for noise-characterization measurements. After calibrating out the electrical-link losses, the measured noise power spectral density of the analog optical link reached −146.69 dBm/Hz. Third-order nonlinear performance was also evaluated on the same setup, yielding an output third-order intercept point (OIP3) above 30 dBm. Using the standard SFDR relation SFDR = 2/3 • (OIP3 − N_floor), the resulting spurious-free dynamic range (SFDR) is approximately 90 dB·$Hz^{2/3}$. These measured results confirm that the proposed A-RoF architecture can simultaneously deliver good noise performance and linearity for 6G wideband fronthaul applications, providing experimental support for the engineering deployment of the analog optical fronthaul domain in Omni-PBS.

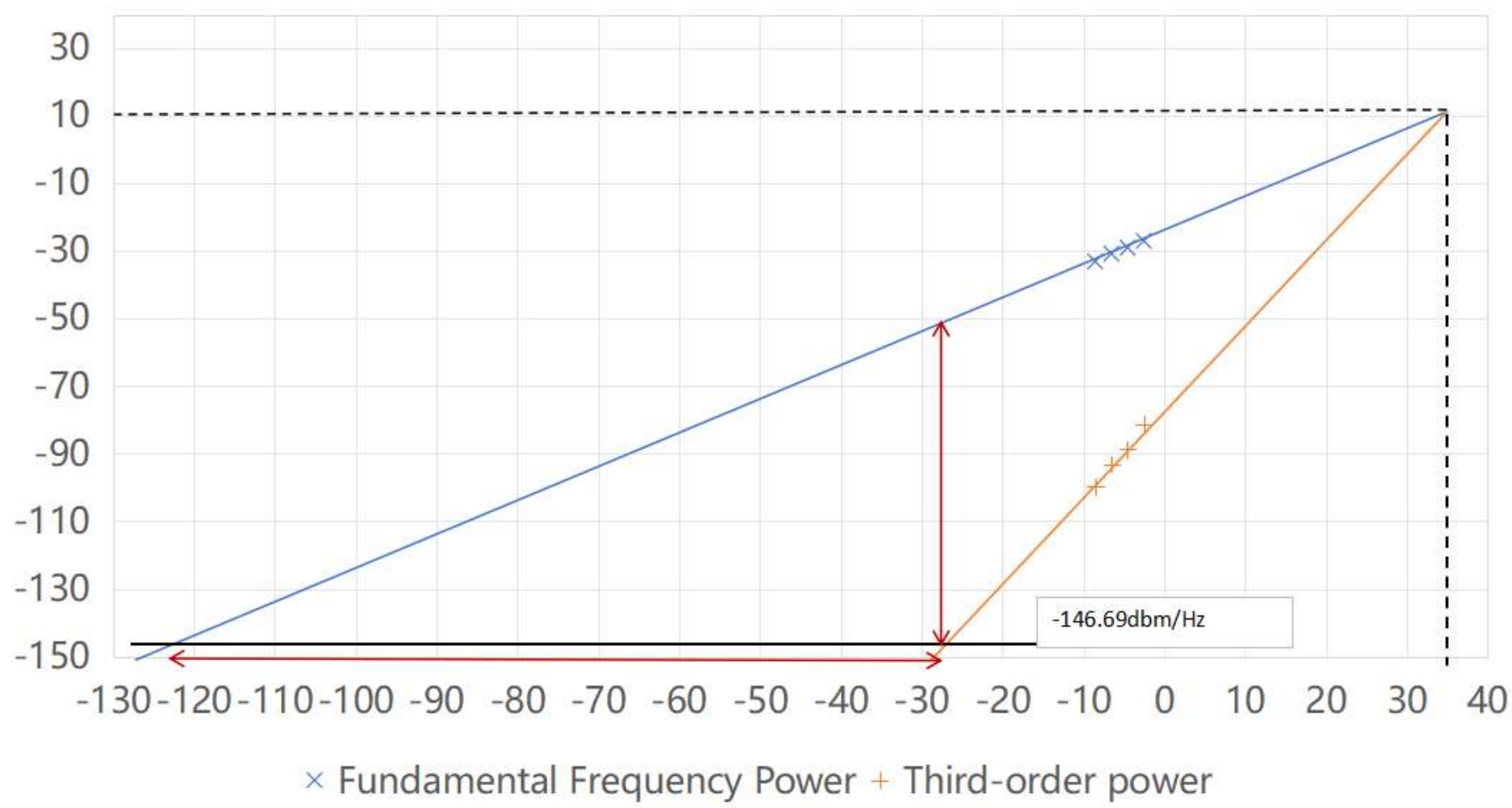

*Fig. 4.6 Measured SFDR of the A-RoF experimental link*

In the A-RoF architecture, the RF waveform is no longer sampled and quantized at the remote end, but is directly modulated onto the optical carrier and transmitted over fiber in analog form. Under this architecture, the structure of the remote unit is greatly simplified: a photodetector (PD) directly recovers the optical signal into an RF electrical signal, which is fed to the antenna after amplification by a power amplifier (PA); no high-speed ADC/DAC or digital baseband processor is required in the link.

The core challenge of A-RoF lies in the nonlinear distortion and noise accumulation of the analog optical link. The relative intensity noise (RIN) of the laser, the nonlinearity of the modulator (IM2/IM3 intermodulation products), and the power fading caused by fiber dispersion together limit the dynamic range of the link [8]. The main factors limiting the link dynamic range include:

1. Laser relative intensity noise (RIN): the inherent intensity fluctuation of the laser is directly superimposed on the RF signal and becomes the main bottleneck of the signal-to-noise ratio at low received optical power;
2. Modulator nonlinearity: the sinusoidal transfer function of the Mach-Zehnder modulator (MZM) introduces second-order intermodulation (IM2) and third-order intermodulation (IM3) products; under multi-carrier scenarios the intermodulation components fall within the signal band and degrade the EVM;
3. Fiber-dispersion-induced power fading: dispersion causes periodic power notches in double-sideband signals at specific frequencies, limiting the available link bandwidth and transmission distance.

The above factors together limit the spurious-free dynamic range (SFDR) of the link to about 100–110 dB·Hz$^{2/3}$. Zhang et al. [8] surveyed three types of linearization techniques: (1) electrical-domain analog predistortion—compensating for modulator nonlinearity at the transmitter to suppress odd-order intermodulation products; (2) optical-domain linearization—using hybrid polarization, dual wavelength, and optical channelization to simultaneously suppress even- and odd-order nonlinear distortion in the optical domain; (3) digital predistortion (DPD)—linearizing the link through baseband signal preprocessing. The survey points out that analog and optical-domain linearization can achieve more than 10 dB of SFDR improvement with bandwidth up to tens of GHz, making them more suitable for wideband fronthaul scenarios; traditional DPD, limited by the bandwidth of analog-to-digital converters, still warrants attention after the introduction of simplified models such as Hammerstein. Under multi-carrier and multi-band multiplexing scenarios, the cross-modulation products between different frequency bands remain a challenge that has not been fully resolved.

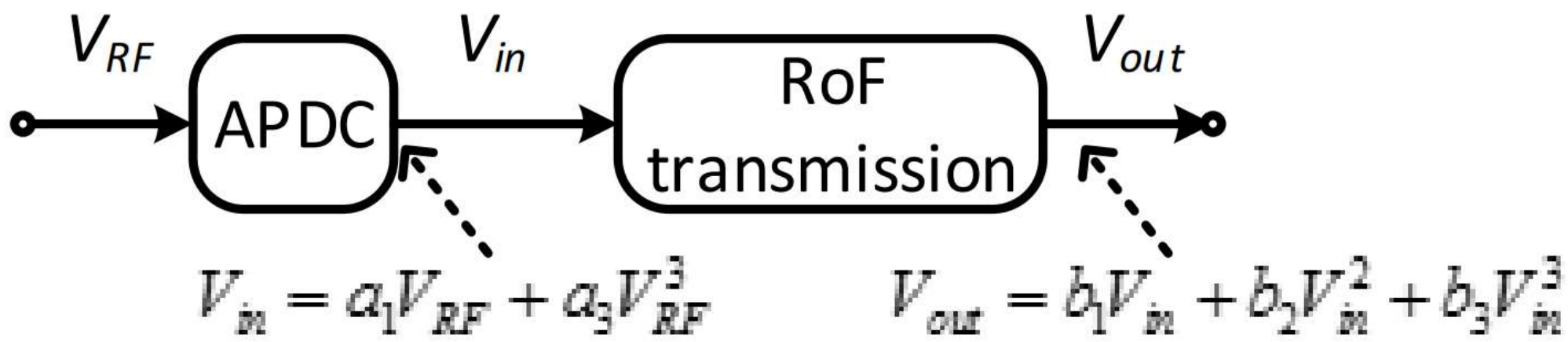

*Fig. 4.7 Schematic of RoF transmission linearization using an analog predistortion circuit (APDC) (after [8] Zhang et al.)*

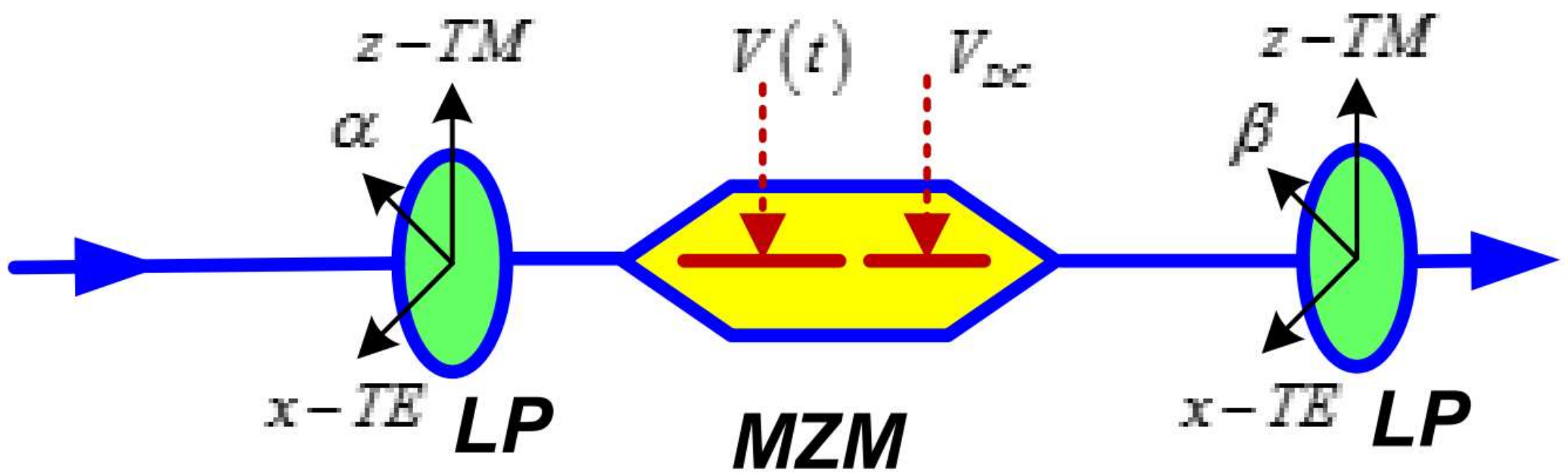

*Fig. 4.8 Hybrid-polarization optical-domain linearization configuration (after [8] Zhang et al.)*

## 5 Microwave-Photonic RF Domain

### 5.1 Physical Bottlenecks of RF Front-Ends at High Frequency Bands

The continuous extension of 6G and its subsequent evolution into millimeter-wave and terahertz bands gradually reveals the device physical limits of traditional electronic RF front-ends. In the millimeter-wave band (28–100 GHz), although GaAs and GaN power amplifiers can provide sufficient gain, their efficiency drops sharply with frequency under wideband operation: the typical efficiency at 28 GHz is about 30%–40%(continuous wave), falling below 15% above 100 GHz [9]. At the same time, electronic phase shifters exhibit frequency-dependent phase shifts under wideband signals, causing the beam direction to skew with frequency (beam squint); the wider the instantaneous bandwidth, the more severe the squint. For the several-GHz instantaneous bandwidth expected by 6G, electronic phase-shifter beam squint can reach several to more than ten degrees, exceeding the beam-alignment tolerance defined by 3GPP [1]. The terahertz band (100–300 GHz) presents even more severe challenges: the cutoff frequency of electronic devices approaches physical limits, the available gain and noise performance of low-noise amplifiers and mixers deteriorate sharply, traditional electronic LOs cannot meet the phase-noise requirements of coherent communications at terahertz frequencies, and the generation and distribution of high-frequency LOs consume extremely high power. As the frequency band rises, the increase in bandwidth also significantly raises the pressure on analog-to-digital conversion and digital processing, especially for latency-critical scenarios, which further raises the cost and power consumption requirements of the base station.

Microwave photonics provides a fundamentally different technical path to break through the above bottlenecks. Capmany and Novak [10] systematically expounded the core idea of microwave photonics in a seminal Nature Photonics review: using photonic technology to generate, transmit, and process microwave signals, introducing the large bandwidth, low loss, and electromagnetic-interference immunity of photons into the microwave domain. Yao [35] further summarized the four functional pillars of microwave photonics—photonic microwave signal generation, photonic microwave filtering, photonic beamforming, and photonic frequency conversion—establishing the functional framework for the photonization of the RF front-end. After more than a decade of development, microwave photonics has moved from laboratory proof-of-principle to engineering applications; the fully photonic coherent radar system demonstrated by Ghelfi et al. [11] is a milestone from device-level to system-level in this field, verifying the complete integration feasibility of photonic RF generation, beamforming, and down-conversion within a single system.

The operating frequency of photonic devices does not depend on carrier transit time but is determined by the optical carrier frequency (~200 THz); therefore, the bandwidth of a microwave-photonic link does not decrease as the RF frequency rises [10], [35]. This physical property enables the photonic RF front-end not only to replace electronic functions at high frequencies, but also to realize functions that electronic devices fundamentally cannot. Table 3 compares the differences between electronic and photonic RF front-ends on key 6G metrics.

*Table 3: Comparison between electronic and photonic RF front-ends*

| Metric | Electronic RF front-end | Photonic RF front-end | Source of advantage |
|---|---|---|---|
| Operating band | Limited by $f_T$ (<300 GHz) | No upper frequency limit (optical carrier ~200 THz) | Optical frequency far above microwave |
| Wideband beamforming | Beam squint (electronic phase shifter) | No squint (true-time delay) | Physical optical-path delay is frequency-independent |
| LO phase noise | Degrades with frequency | Determined by optical comb (low and stable) | Optical comb frequency stability |
| Multi-channel LO distribution | Large loss in electrical distribution network | Lossless distribution by optical splitter | Low fiber loss |
| Instantaneous bandwidth | Limited by ADC/DAC sampling rate | Several GHz (photonic bandwidth) | No high-speed ADC/DAC required |
| Electromagnetic interference | Sensitive | Immune | Photons are non-electromagnetically coupled |

The following subsections analyze the four functions—beamforming, filtering and interference identification, mixer down-conversion, and terahertz generation—in turn.

## 5.2 Functional Implementation of the Photonic RF Front-End

With the continuous rise in wireless communication frequencies and the rapid increase in bandwidth, the beamforming scheme for massive antenna arrays is gradually evolving from fully digital beamforming toward hybrid analog-digital beamforming, in which the share of analog beamforming keeps growing. Optical true-time-delay (TTD) analog beamforming is the core function of photonization in the RF domain. TTD steers the beam by controlling the transmission delay of the signal in each antenna channel. Because the delay produces the same phase shift for all frequency components of a wideband signal ($\varphi = 2\pi f\tau$, with $\tau$ constant), TTD fundamentally eliminates beam squint. The fully photonic coherent radar system published by Ghelfi et al. [11] in Nature is a landmark work in this direction. The system fully implements the RF signal generation, receive beamforming, and down-conversion chain in the optical domain, verifying TTD beamforming capability in the X-band (8–12 GHz) with beam-pointing accuracy better than 0.5° and no beam squint within a 4 GHz instantaneous bandwidth. This achievement not only proves the system-level feasibility of the photonic RF front-end, but also establishes the technical paradigm of "optical-domain continuity"—completing signal generation to processing entirely in the optical domain, with optoelectronic conversion only at the final radiation and initial reception.

Reconfigurable filtering capability of the RF front-end is crucial for the dynamic spectrum environment of 6G. Traditional electronic filters (SAW/BAW) have fixed bandwidth, and reconfiguration requires switching physical filters at millisecond speed. Yao [35] systematically expounded the theoretical framework of photonic microwave filtering in a microwave-photonics review: the RF signal is converted to the optical domain by electro-optic modulation, and through an optical delay-line network (or dispersive medium) tap delays are produced; each tap is weighted by a variable optical attenuator and summed, and the filtered RF signal is recovered by photodetection. By adjusting the number of taps, the delay interval, and the weight coefficients, FIR or IIR filter responses can be realized. Building on this theory, [36] used two cascaded high-Q microrings to achieve a bandpass filter with center frequency 0.84–41.06 GHz, bandwidth 0.84–18.61 GHz tunable, and out-of-band suppression up to 35 dB. Compared with the millisecond-level reconfiguration of electronic filters, the microsecond-level speed of photonic filters realizes truly "dynamic" filtering.

The physical basis of photonic frequency conversion is optical heterodyne beat: two optical waves whose frequency difference equals the target RF frequency produce a difference-frequency component in a high-speed photodetector through square-law detection, realizing frequency translation from the optical domain to the microwave/millimeter-wave/terahertz domain [37]. A unique advantage of the optical heterodyne scheme is the use of multiple longitudinal modes of the same optical comb to simultaneously generate carriers at different frequencies. An optical comb with a 25 GHz mode spacing can simultaneously generate multiple carrier frequencies from 25 GHz to 300 GHz; each mode pair is independently beat after WDM separation [38]. This capability enables the Omni-PBS to support Sub-6GHz, millimeter-wave, and terahertz multi-band operation simultaneously with a single optical comb source, without configuring a separate electronic oscillator for each band—a capability of practical significance for the large-scale multi-band coverage of 6G ubiquitous-connectivity scenarios in terms of power consumption and complexity.

In the uplink, the wideband RF signal received by the antenna must be down-converted to an intermediate or baseband frequency for subsequent processing. Photonic down-conversion uses the symmetric optical heterodyne principle of up-conversion: the received RF signal is converted to the optical domain by electro-optic modulation and then beat with the optical LO, producing a difference-frequency (intermediate-frequency) signal in the photodetector [39]. Its instantaneous bandwidth also exceeds 40 GHz, and its LO leakage is 10–20 dB lower than that of electronic mixers. Gao et al. [40] further cascaded a photonic mixer with a programmable photonic filter, simultaneously completing down-conversion and filtering in the optical domain, integrating the three discrete modules of mixer, LO, and filter in traditional receiver chains onto a single photonic platform. The filter directly suppresses adjacent-channel interference after down-conversion, avoiding the electronic filtering bottleneck at the intermediate-frequency stage.

### 5.3 Photonic RF Front-End Integration Scheme

The previous section discussed the four core functions of photonic TTD beamforming, programmable microwave-photonic filtering, photonic frequency conversion, and channelized reception. Integrating the above photonic RF front-end functions into a chip-scale photonic RF front-end deployable in base stations requires solving three integration challenges: multi-material-platform compatibility, packaging technology, and thermal management.

The silicon-photonics platform (SOI-based) is the optimal substrate for photonic RF front-end integration. Its manufacturing process is CMOS-compatible, enabling cost reduction through mature 12-inch wafer production lines; SOI waveguides have loss below 1 dB/cm in the C-band (1550 nm), naturally matching the operating band of A-RoF fronthaul and WDM multiplexing systems. The silicon-photonics platform has verified the integration of core devices required for the photonic RF front-end—AWG wavelength multiplexing/demultiplexing, Mach-Zehnder interferometers, microring resonators, and thermo-optic tunable delay lines [35]—with a layout density sufficient to implement a 256-channel TTD network within a chip area of about 10 mm×10 mm [11]. However, silicon photonics alone cannot meet all requirements with a single material: high-speed photodetection (>100 GHz bandwidth) requires InP-based UTC-PD or SiGe detectors, low-loss high-speed electro-optic modulation requires thin-film lithium niobate (TFLN), and high-purity optical comb sources require SiN microrings or InP gain chips. Therefore, the integration scheme of the photonic RF front-end must inevitably be heterogeneous integration.

This paper proposes a three-layer packaging scheme of "silicon-photonics substrate + multi-material heterogeneous integration." The first layer is the silicon-photonics functional layer, monolithically processed on an SOI wafer, integrating devices that need no special material—AWG, MZI tunable delay lines, microring filters, thermo-optic variable optical attenuators (VOA), and optical waveguide interconnects. The second layer is heterogeneous-material functional devices, integrated onto the silicon-photonics substrate through wafer bonding or chip attachment. The TFLN modulator achieves high-speed electro-optic modulation through heterogeneous integration of a BTO thin film with a silicon waveguide [41], used for RF signal modulation in photonic frequency conversion. InP-based photodetectors can be flip-chip bonded above the grating coupler of the silicon-photonics chip [42], used for photodetection and terahertz generation. Ge-Si photodetectors can be directly integrated in the SOI process for non-high-speed detection channels. The third layer is the electronic driving and control layer, interconnected with the photonic chip through through-silicon vias (TSV) or flip-chip, containing the thermo-optic driving DAC of the TTD network, the control circuit of the VOA, the transimpedance amplifier (TIA) of the UTC-PD, and the phase-locked control circuit of the optical comb source, all implemented in CMOS technology.

The optical comb source is the core of photonic LO distribution and terahertz generation, and also the most difficult module to integrate. The mode-locked laser (MLL) requires an InP gain chip plus an external cavity, is bulky and difficult to integrate on-chip; the electro-optic modulation comb requires high-frequency RF driving and high power consumption. The microresonator Kerr comb generates an optical comb using the nonlinear effect in SiN microrings and can be integrated on the silicon-photonics platform, but requires a high-Q microring ($Q>10^6$) and precise pump-laser coupling. Kippenberg et al. [38] systematically reviewed the theoretical foundation and experimental progress of microresonator Kerr combs; SiN microresonator Kerr combs have achieved comb output covering an 80 nm range in the C-band with linewidth below 1 kHz. The currently most feasible integration scheme is hybrid integration of an SiN microresonator Kerr comb chip with the silicon-photonics functional layer through edge coupling or grating coupling, with the pump laser provided by an III-V gain chip.

In packaging, both optical alignment accuracy and RF interconnection integrity must be satisfied. Co-packaged optics (CPO) technology realizes low-loss optical interconnection through passive alignment of photonic and electronic chips; RF interconnection uses flip-chip bumps or TSVs, and bump pitch below 50 μm makes RF parasitics negligible [42]. In thermal management, a TEC combined with a heat sink can control chip temperature fluctuation within ±0.1°C, meeting the thermal stability requirements of MZIs and microrings. The remote unit must keep the chip temperature stable in an ambient temperature range of −40°C to +55°C, a requirement within the experience range of current base-station RF unit thermal design.

The chip-scale integration of the photonic RF front-end advances in three stages. The first stage adopts a hybrid integration scheme, where the silicon-photonics chip, InP detector chip, and TFLN modulator chip are interconnected by fiber jumpers and packaged as a multi-chip module (MCM); the integration density is low but the realizability is high, suitable for functional verification and field trials. The second stage transitions to 2.5D heterogeneous integration, where the material chips are interconnected through silicon-photonics waveguides on an interposer, eliminating fiber jumpers and improving packaging density by an order of magnitude. The third stage realizes 3D monolithic heterogeneous integration, where TFLN thin films and InP materials are directly epitaxially grown or wafer-bonded onto the SOI substrate, and all photonic devices are implemented on a single chip. The yield of heterogeneous-integration processes is the key bottleneck for the system integration transition of chip systems; it is expected to gradually reach economic feasibility as processes are standardized and scale effects emerge.

# 6 System-Level Co-Design

## 6.1 Cross-Domain Co-Design Principles

The end-to-end optical-domain continuity principle requires that the three domains jointly consider the global distribution of optoelectronic conversion boundaries in signal-chain design, rather than independently optimizing each domain's own conversion overhead. The output interface of the baseband signal from the optical computing domain must match the modulation scheme of the fronthaul domain, so that the fronthaul domain can directly carry this signal for A-RoF transmission. The interface between the fronthaul and RF domains similarly requires the optical-carrier parameters and functional-domain requirements of A-RoF to match the operating band and interface specifications of the photonic RF front-end—for example, generating coherent light sources required by photonic frequency conversion must be provided at the fronthaul stage. Consistency of cross-domain parameters and requirements is the physical precondition for optical-domain continuity, and is also the fundamental characteristic that distinguishes the unified Omni-PBS architecture from deploying photonic technology independently in the three domains.

The power-budget coordination principle of remote-unit simplification and RF photonization manifests as a positive feedback relationship between the fronthaul and RF domains. After A-RoF eliminates remote ADCs/DACs and digital processing, the power budget and physical space of the remote end are released, creating conditions for introducing photonic beamforming and photonic filtering devices in the RF domain. Conversely, if the remote end still retains digital processing circuits, superimposing a photonic RF front-end on this basis would cause the total remote-end power consumption to exceed thermal constraints, making RF photonization unable to meet current specification requirements. Therefore, the introduction order of the two domains has a certain dependency: the A-RoF simplification of the fronthaul domain should precede or be synchronous with the photonization deployment of the RF domain. It should be emphasized that, by virtue of the architectural advantages of A-RoF, the centralized pooling of baseband and intermediate-frequency circuits, in conjunction with the substantial simplification of remote units, yields greater cost and energy reductions for the entire base station, even though RF photonics introduces additional power consumption.

The joint-optimization principle of functional splitting couples the Split choice of the fronthaul domain with the functional allocation of the baseband and RF domains. Low-layer splitting (Split 7.2x and below) retains more baseband functions on the BBU side, and the advantages of the digital fronthaul solution are most pronounced. Although this reduces fronthaul bandwidth, it increases complexity at the remote end, which runs counter to the goal of A-RoF to simplify the remote end. The Omni-PBS recommends using a segmentation point higher than that of Option 8 as the baseline segmentation point; under these conditions, the efficiency of three-domain coordination is optimized.: high baseband computing concentration (photonic accelerators can be deployed at scale), fronthaul signal form adapted to A-RoF (analog RF signal directly optically modulated for transmission), and no digital processing required at the remote end (RF-domain photonization feasible within the power budget).

## 6.2 Processor-Decoupled Optical Computing Architecture

In traditional optical computing systems, the control part is usually handled by an electronic processor, forming an optoelectronic hybrid computing architecture. The electronic processor is responsible for data movement, instruction issuance, and heterogeneous computing resource scheduling during computation, as shown in the red section in the Figure 6.1. However, due to the characteristics of optical computing, the bus, interface, and processor resources are occupied for a long time during computation, resulting in resource waste. The processor-decoupled optical computing architecture can sink task scheduling, data transformation, and computation-instruction generation entirely to the storage side, leaving the processor only in the lightweight role of initiating the flow and reclaiming results, while improving optical computing utilization.

The processor-decoupled optical computing architecture consists of three parts: the processor, the intelligent memory, and the optical computing unit, as shown in the orange section in the figure 6.. At the task-initiation stage, the processor performs heterogeneous computing resource scheduling according to the computing requirements of the upper-layer application, determines the optical computing scale and data source, and completes preprocessing operations such as precision alignment and data blocking on the original data. The processor then sends the data to be processed together with structured control information to the intelligent memory. The control information explicitly defines the optical-computing start identifier, the required number of computations, and the task priority, thereby granting the storage side complete execution autonomy.

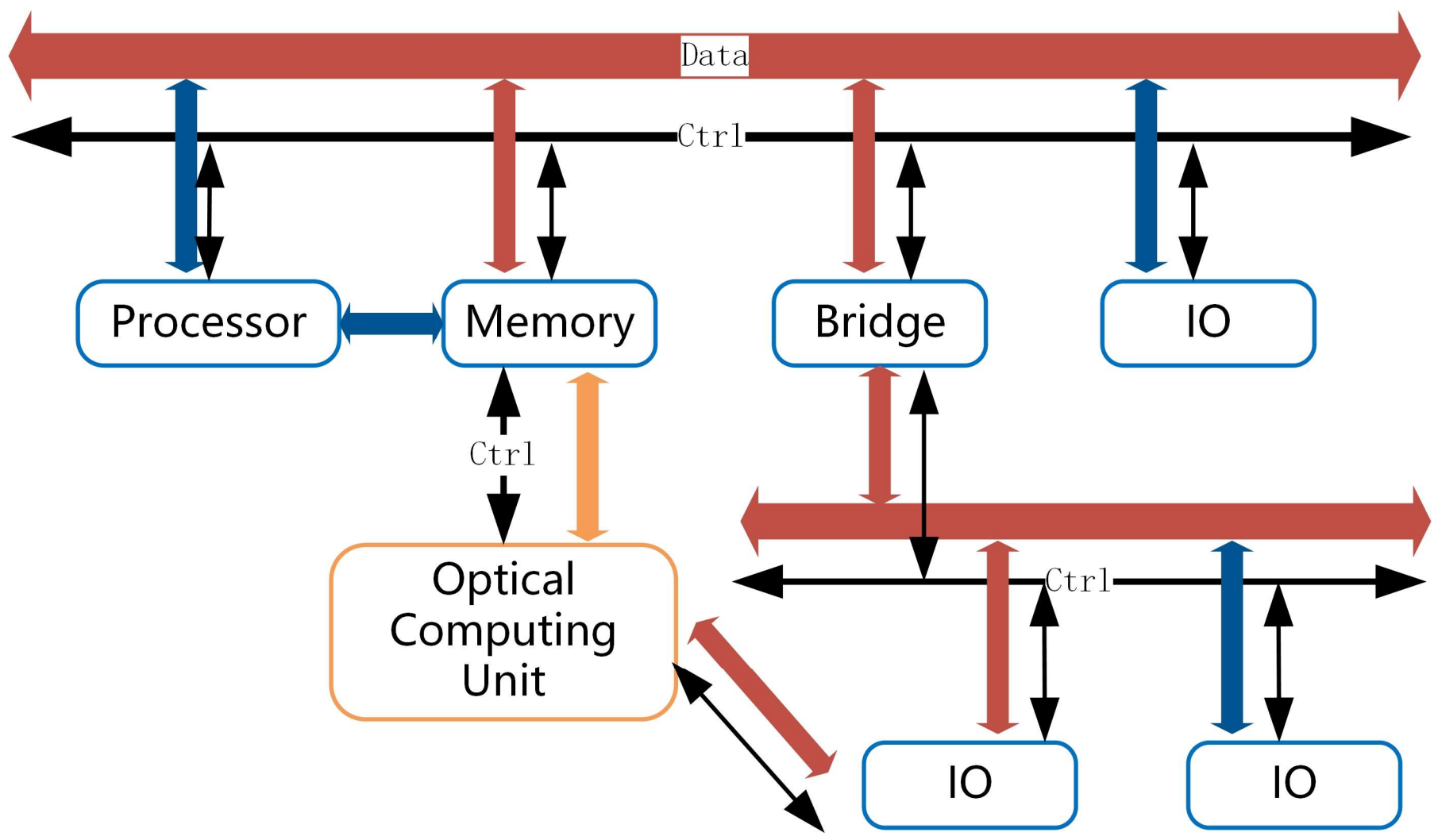

*Fig. 6.1 Processor-decoupled optical computing architecture*

The intelligent memory is responsible for storing, controlling optical computing tasks, and electro-optic/photoelectric conversion. After parsing the control information, the control unit does not immediately initiate computation; instead, it first sends a computing-resource request instruction to the target optical computing unit, requesting the allocation of computing resources. The optical computing unit arbitrates based on its current working state and the priority contained in the request, and returns a resource-request reply. The control unit generates a channel-establishment instruction based on this reply to build a bidirectional optical transmission channel between the memory and the optical computing unit; the memory starts high-speed optical transmission of the pending data after confirming the configuration is complete. This resource-negotiation mechanism enables high-priority tasks to obtain low-latency resource guarantees, entirely independent of the processor. The optical computing unit directly performs optical computing operations on the transmitted pending data, obtains the optical computing result, and returns it directly to the pre-allocated storage space, while the storage control reports task completion to the processor. Throughout the optical computing execution, the processor's timing, interface, and bus resources are completely released and can concurrently process other tasks, thereby significantly improving the system's energy-efficiency performance.

## 7 Discussion: Challenges and Opportunities

### 7.1 Engineering Constraints on Optical Computing Precision and Reliability

The analog-domain precision of the photonic computing accelerators in the optical baseband computing domain is the primary constraint for the Omni-PBS toward real deployment. The equivalent precision of MZI arrays is about 6–8 bit ENOB, and that of $D^2NN$ is about 4–6 bit [2], [12], both lower than the standard FP32 of electronic processors. The precision of analog optical computing is affected by device manufacturing consistency, temperature drift, and aging; the equivalent precision of the same MZI array may fluctuate by 1–2 bit under different operating conditions, leading to time-varying inference performance.

The solution path includes three levels. At the device level, the thin-film lithium niobate (TFLN) platform can provide higher modulation linearity and thermal stability than silicon-based MZIs, promising to raise the equivalent precision to 8–10 bit [43]. At the algorithm level, quantization-aware training (QAT) and noise-aware training (NAT) inject equivalent-noise and quantization constraints during the training stage, enabling the model to converge to a better solution under the target precision [2]. At the system level, critical tasks (such as channel decoding) retain electronic processing, and only tasks with larger tolerance are offloaded to photonic accelerators, forming a mixed-precision scheduling strategy.

### 7.2 Linearization and Dynamic Range of the Analog Optical Link

Analog optical fronthaul (Analog Radio-over-Fiber, A-RoF), as a key technology for implementing centralized radio access networks (C-RAN) in 5G and future 6G mobile communication networks, has its core advantage in extremely high spectral efficiency and a minimal remote radio unit (RRH) architecture [44]. At the device level, insufficient output power of the photodetector in the fronthaul link is one of the key bottlenecks restricting A-RoF remote-end simplification. In traditional IM-DD analog optical links, the remote end uses a PIN photodetector (PIN-PD) to recover the optical signal into an RF electrical signal. Limited by the space-charge effect and carrier transit time of the PIN-PD, its saturated optical power is typically only 0 to +10 dBm; under typical fronthaul optical-power budgets the output RF power is only −20 to −10 dBm [32], and the output power capability of the detector itself becomes the key physical constraint determining the simplification degree of the RRU architecture. The uni-traveling-carrier photodiode (UTC-PD) separates the optical absorption layer from the carrier collection layer, using only electrons as active carriers and avoiding the slow transit-time limitation of holes, thereby achieving a saturated optical power far higher than that of PIN-PD—the current UTC-PD can reach a saturated input optical power of +20 dBm or above, with output RF power at the +15 to +20 dBm level [32], [33]. Vega et al. [34] demonstrated a compact Ka-band transmitter scheme with UTC-PD directly connected to the antenna, verifying the feasibility of the "PD-direct-drive-antenna" architecture in the millimeter-wave band. Deploying UTC-PD at the RRU end is expected to fundamentally eliminate the PA and its linearization circuit, reducing the types of active devices in the remote-end transmit chain from more than ten in the traditional scheme to only the PD, and is one of the core enabling technologies for realizing the "minimalization" of the Omni-PBS RRU.

At the link level, the performance of A-RoF is fundamentally limited by the nonlinear distortion and dynamic range of the analog optical link, which directly affects the error vector magnitude (EVM) performance under high-order modulation formats (such as 64-QAM) [45]. Although compensation techniques for major impairments—modulator nonlinearity, relative intensity noise (RIN), and dispersive power fading—are maturing at the theoretical level, they still face significant challenges in engineering deployment, especially in complex multi-carrier and multi-channel multiplexing scenarios [46]. In terms of dynamic range and linearity, the 5G New Radio (NR) orthogonal frequency-division multiplexing (OFDM) signal has a peak-to-average power ratio (PAPR) as high as 8–10 dB, imposing stringent requirements on the spurious-free dynamic range (SFDR) of the link. Studies show that to meet the EVM indicator of 64-QAM modulation, the A-RoF link requires an SFDR greater than 110 $dB\cdot Hz^{2/3}$; DPD has become the mainstream scheme for improving linearity, and can improve link SFDR by 5–8 dB by compensating for memory and memoryless nonlinearities [47]. However, the introduction of DPD weakens to some extent the core advantage of A-RoF that "the remote end has no digital processing"; although DPD coefficient calculation is concentrated on the BBU side rather than the remote end, it significantly increases the computing load on the BBU side [48]. When the BBU side simultaneously runs physical-layer optical computing tasks, a strict resource-coordination mechanism must be established to guarantee the real-time performance of DPD updates; otherwise, lagging nonlinearity compensation will cause EVM degradation [49]. Given that supporting a DPD feedback link at the remote unit would increase its complexity and cost, open-loop or semi-open-loop DPD can be considered to correct the nonlinearity of the remote unit, which is expected to satisfy the 3GPP pico base station specifications with relatively low computational and hardware complexity.Recent studies indicate that machine-learning-based DPD schemes (such as the augmented real-valued time-delay neural network, ARVTDNN) outperform traditional generalized memory polynomial (GMP) methods in handling memory nonlinearity, but also bring higher algorithm complexity [47].

In ultra-wideband multi-band and multi-channel multiplexing scenarios, In the first stage, after the introduction of photonics, when large-bandwidth multi-channel signals share a single fiber, the dynamic range at each stage of the optical link may fall short—manifesting as insufficient linearity and elevated noise—given a PAPR of 15 dB and an SNR above 30 dB. In the second stage, the dynamic-range problem is particularly prominent. When wavelength-division multiplexing (WDM-A-RoF) technology is used to multiplex up to 64 antenna channels onto the same fiber, inter-channel crosstalk and gain non-flatness of optical amplifiers become the main bottlenecks [50]. To ensure that the EVM of each channel meets the standard, crosstalk between adjacent channels must be suppressed below −30 dBc [51]. However, the gain flatness of commercial erbium-doped fiber amplifiers (EDFA) is typically only 1–2 dB, which causes significant channel-power differences in multi-wavelength systems [52]. To overcome this limitation, variable optical attenuators (VOA) must be introduced for channel equalization, which not only increases the number of photonic devices but also raises the complexity and cost of the remote node, partially offsetting the original intention of simplifying the remote-end architecture with A-RoF [53]. In addition, the four-wave mixing (FWM) nonlinear effect in WDM systems further exacerbates inter-channel interference under high-power injection, which must be suppressed by optimizing channel spacing and power management [46].

Overall, the engineering implementation of the Omni-PBS analog optical fronthaul link needs to advance simultaneously along three dimensions: at the device dimension, replacing traditional PIN-PD with high-power UTC-PD to eliminate the remote-end PA and its linearization circuit, realizing the structural simplification of the RRU architecture; at the link dimension, guaranteeing nonlinear-distortion and dynamic-range indicators through DPD, adaptive filtering, and modulation-depth optimization; at the system dimension, decoupling the traditional coupling between fronthaul-bandwidth growth and remote-end complexity growth through WDM multi-channel parallel transmission and BBU centralized processing, providing a scalable fronthaul solution for the ultra-dense deployment of 6G ultra-massive antenna arrays.

### 7.3 Integration Density and Cost of the Photonic RF Front-End

The engineering of the photonic RF front-end faces dual challenges of integration density and cost. The photonic TTD beamforming network contains a large number of tunable delay units and optical switches; a 256-element array requires 256 independent delay channels, occupying a chip area of about 10 mm×10 mm on the silicon-photonics platform [11]. The current silicon-photonics tape-out cost is about $50,000 per batch, with a per-die cost of about $500–1,000; compared with electronic RF front-end chips (about $50–100 per channel), it does not yet have a cost advantage.

In terms of integration density, the functional modules of the photonic RF front-end—including the TTD network, photonic filters, photonic mixers, and optical comb sources—are currently implemented in discrete or hybrid integration, and no monolithic integration scheme exists yet. The silicon-photonics platform can integrate TTDs and filters, but optical comb sources and high-speed photodetectors usually require InP or TFLN materials; the packaging yield and reliability of heterogeneous integration (SiPh + InP + TFLN) still need verification [54]. Table 4 estimates the integration maturity of each module of the photonic RF front-end.

*Table 4: Integration-maturity assessment of photonic RF front-end modules*

| Module | Current integration approach | Monolithic-integration feasibility | Maturity |
|---|---|---|---|
| Photonic TTD network | Silicon-photonic integration | Demonstrated (<64 elements) | Medium |
| Photonic microwave filter | Silicon-photonic integration | Demonstrated | Medium |
| Photonic mixer | Discrete / hybrid integration | Feasible on silicon photonics (requires high-speed PD) | Low–Medium |
| Optical comb source | InP / nonlinear fiber | Requires heterogeneous integration | Low |
| WDM multiplexer / demultiplexer | Silicon-photonic AWG | Demonstrated | High |
| High-speed photodetector | InP / UTC-PD | Requires heterogeneous integration | Medium |

The cost-reduction path depends on the scaling of silicon-photonics tape-outs and the maturation of heterogeneous-integration processes. Referring to the cost-decline curve of optical transceiver modules (100G CWDM4 modules fell from $2,000 in 2016 to $150 in 2023), the photonic RF front-end is expected to reach a cost level comparable to that of electronic RF front-ends around 2030.

### 7.4 Natural Fit Between New 6G Scenarios and Photonic Technologies

Although the above challenges are severe, the demands of the three major 6G scenarios on optical technologies are irreplaceable, constituting the fundamental driving force for the development of the Omni-PBS.

The integrated sensing and communication (ISAC) scenario requires the base station to perform communication and sensing functions simultaneously. The sensing function requires an RF front-end with large instantaneous bandwidth and high dynamic range; the bandwidth of photonic microwave links is not limited by the RF frequency (Table 3), and the photonic channelized receiver can process the communication and sensing bands simultaneously [35]. In addition, photonic TTD beamforming supports the simultaneous formation of communication beams and sensing beams without time-division switching, naturally matching the continuous-sensing requirement of ISAC.

The AI-native air interface (AIAC) scenario requires real-time AI inference at the baseband. The microsecond-level latency of GPUs can hardly satisfy the latency constraints of physical-layer AI inference, and the picosecond-level latency of photonic accelerators provides the feasible solution [3], [12]. 3GPP positions AI/ML as a KPI enhancement rather than a baseline function, which means that the precision loss of optical computing is within an acceptable range, while its latency and energy-efficiency advantages are of decisive significance.

The ultra-dense connectivity (UC) scenario requires the large-scale deployment of distributed remote units. The minimalist remote end simplified by A-RoF (<5 W, no digital circuits) makes the dense deployment of thousands of RRUs feasible in terms of both power consumption and cost [52]. Photonic LO distribution enables coherent cooperation among multiple RRUs, providing the hardware foundation for distributed MIMO and cooperative communication [11].

The superposition effect of the three scenarios makes the Omni-PBS not an "optional optimization" but a "necessary architecture." A single scenario might barely be handled by electronic technology, but when the three scenarios coexist, the triple bottlenecks of the electronic base station will erupt simultaneously, and the systematic introduction of optical technology becomes the only way out.

## References

[1] "Study on Artificial Intelligence (AI)/Machine Learning (ML) for NR Air Interface," 3GPP, TR 38.812, Rel. 18, 2024. [Online]. Available: https://www.3gpp.org/technologies/ai-ml-nr. [Accessed: Jul. 15, 2026].

[2] G. Wetzstein *et al.*, "Inference in artificial intelligence with deep optics and photonics," *Nature*, vol. 588, no. 7836, pp. 39–47, Dec. 2020, doi: 10.1038/s41586-020-2973-6.

[3] M. A. Nahmias, T. F. De Lima, A. N. Tait, H.-T. Peng, B. J. Shastri, and P. R. Prucnal, "Photonic Multiply-Accumulate Operations for Neural Networks," *IEEE J. Select. Topics Quantum Electron.*, vol. 26, no. 1, pp. 1–18, Jan. 2020, doi: 10.1109/JSTQE.2019.2941485.

[4] J. Feldmann *et al.*, "Parallel convolutional processing using an integrated photonic tensor core," *Nature*, vol. 589, no. 7840, pp. 52–58, Jan. 2021, doi: 10.1038/s41586-020-03070-1.

[5] "CPRI Specification V7.0," CPRI Cooperation, 2015.

[6] P. T. Dat, A. Kanno, N. Yamamoto, and T. Kawanishi, "Seamless Convergence of Fiber and Wireless Systems for 5G and Beyond Networks," *Journal of Lightwave Technology*, vol. 37, no. 2, pp. 592–605, Jan. 2019, doi: 10.1109/JLT.2018.2883337.

[7] C. Lim and A. Nirmalathas, "Radio-Over-Fiber Technology: Present and Future," *Journal of Lightwave Technology*, vol. 39, no. 4, pp. 881–888, Feb. 2021, doi: 10.1109/JLT.2020.3024916.

[8] X. Zhang, R. Zhu, D. Shen, and T. Liu, "Linearization Technologies for Broadband Radio-Over-Fiber Transmission Systems," *Photonics*, vol. 1, no. 4, pp. 455–472, Dec. 2014, doi: 10.3390/photonics1040455.

[9] T. S. Rappaport *et al.*, "Wireless Communications and Applications Above 100 GHz: Opportunities and Challenges for 6G and Beyond," *IEEE Access*, vol. 7, pp. 78729–78757, 2019, doi: 10.1109/ACCESS.2019.2921522.

[10] J. Capmany and D. Novak, "Microwave photonics combines two worlds," *Nature Photon*, vol. 1, no. 6, pp. 319–330, Jun. 2007, doi: 10.1038/nphoton.2007.89.

[11] P. Ghelfi *et al.*, "A fully photonics-based coherent radar system," *Nature*, vol. 507, no. 7492, pp. 341–345, Mar. 2014, doi: 10.1038/nature13078.

[12] X. Lin *et al.*, "All-optical machine learning using diffractive deep neural networks," *Science*, vol. 361, no. 6406, pp. 1004–1008, Sep. 2018, doi: 10.1126/science.aat8084.

[13] X. Zhao, S. Xu, S. Yi, S. Hua, X. Li, and W. Zou, "Photonic parallel channel estimation of MIMO-OFDM wireless communication systems," *Opt Express*, vol. 31, no. 2, pp. 1394–1408, Jan. 2023, doi: 10.1364/OE.476556.

[14] J. Jin *et al.*, "Adaptive time-delayed photonic reservoir computing based on Kalman-filter training," *Opt Express*, vol. 30, no. 8, pp. 13647–13658, Apr. 2022, doi: 10.1364/oe.454852.

[15] "eCPRI Specification V2.0," O-RAN Alliance, 2019.

[16] D. Che, “Analog vs Digital Radio-Over-Fiber: A Spectral Efficiency Debate From the SNR Perspective,” *Journal of Lightwave Technology*, vol. 39, no. 16, pp. 5325–5335, Aug. 2021, doi: 10.1109/JLT.2021.3102220.
[17] A. Delmade *et al.*, “Optical Heterodyne Analog Radio-Over-Fiber Link for Millimeter-Wave Wireless Systems,” *Journal of Lightwave Technology*, vol. 39, no. 2, pp. 465–474, Jan. 2021, doi: 10.1109/JLT.2020.3032923.
[18] M. U. Hadi, “Mitigation of nonlinearities in analog radio over fiber links using machine learning approach,” *ICT Express*, vol. 7, no. 2, pp. 253–258, Jun. 2021, doi: 10.1016/j.icte.2020.11.002.
[19] D. Marpaung, J. Yao, and J. Capmany, “Integrated microwave photonics,” *Nature Photon*, vol. 13, no. 2, pp. 80–90, Feb. 2019, doi: 10.1038/s41566-018-0310-5.
[20] T. Nagatsuma, G. Ducournau, and C. C. Renaud, “Advances in terahertz communications accelerated by photonics,” *Nature Photon*, vol. 10, no. 6, pp. 371–379, Jun. 2016, doi: 10.1038/nphoton.2016.65.
[21] V. J. Urick, K. J. Williams, and J. D. McKinney, Fundamentals of Microwave Photonics. Hoboken, NJ, USA: John Wiley & Sons, Inc., 2015.
[22] Y. Shen *et al.*, “Deep learning with coherent nanophotonic circuits,” *Nature photonics*, vol. 11, no. 7, pp. 441–446, 2017.
[23] C. He, S. Motooka, and S. Sunada, “Photonic Reservoir Computing for Signal Equalization in Optical Wireless Communications,” *IEEE Photonics Technology Letters*, vol. 38, no. 13, pp. 896–899, Jul. 2026, doi: 10.1109/LPT.2025.3618902.
[24] Z. Yu, Y. Zhao, Z. Li, S. Xu, and W. Zou, “Low-Complexity Photonic Wireless Channel Estimation System Based on Matrix- Condition-Number Theory,” *IEEE Photonics Technology Letters*, vol. 37, no. 6, pp. 333–336, Mar. 2025, doi: 10.1109/LPT.2025.3545867.
[25] A. Khaled, A. Aadhi, C. Huang, A. N. Tait, and B. J. Shastri, “Fully integrated hybrid multimode-multiwavelength photonic processor with picosecond latency,” *Nat Commun*, vol. 17, no. 1, p. 28, Dec. 2025, doi: 10.1038/s41467-025-66561-7.
[26] S. Hua *et al.*, “An integrated large-scale photonic accelerator with ultralow latency,” *Nature*, vol. 640, no. 8058, pp. 361–367, Apr. 2025, doi: 10.1038/s41586-025-08786-6.
[27] Z. Xu, T. Zhou, M. Ma, C. Deng, Q. Dai, and L. Fang, “Large-scale photonic chiplet Taichi empowers 160-TOPS/W artificial general intelligence,” *Science*, vol. 384, no. 6692, pp. 202–209, Apr. 2024, doi: 10.1126/science.adl1203.
[28] J. Wang *et al.*, “Microcomb-enabled parallel self- calibration optical convolution streaming processor,” *Light Sci Appl*, vol. 15, no. 1, p. 149, Mar. 2026, doi: 10.1038/s41377-025-02093-5.
[29] A. Checko *et al.*, “Cloud RAN for Mobile Networks—A Technology Overview,” *IEEE Communications Surveys & Tutorials*, vol. 17, no. 1, pp. 405–426, 2015, doi: 10.1109/COMST.2014.2355255.
[30] M. Sung *et al.*, “Photonic THz Communications Based on Radio-Over-Fiber Technology for 6G Mobile Network: Design and Opportunity,” *IEEE Journal of Selected Topics in Quantum Electronics*, vol. 29, no. 5: Terahertz Photonics, pp. 1–11, Sep. 2023, doi: 10.1109/JSTQE.2023.3308899.
[31] R. H. Walden, “Analog-to-digital converter survey and analysis,” *IEEE Journal on Selected Areas in Communications*, vol. 17, no. 4, pp. 539–550, Apr. 1999, doi: 10.1109/49.761034.
[32] T. Umezawa, S. Nakajima, A. Matsumoto, K. Akahane, and N. Yamamoto, “Ultra-Broadband UTC-PD Using Well-Optimized InGaAs/InP Active Layer Toward 200-GHz Bandwidth and Beyond,” in *2023 Conference on Lasers and Electro-Optics Europe & European Quantum Electronics Conference (CLEO/Europe-EQEC)*, Jun. 2023, pp. 1–1. doi: 10.1109/CLEO/Europe-EQEC57999.2023.10232174.
[33] S. Vega *et al.*, “Direct connection of uni-traveling-carrier photodiodes to antennas for frequency reconfigurable fiber-radio transmission in the Ka band,” *Optics & Laser Technology*, vol. 174, p. 110637, Jul. 2024, doi: 10.1016/j.optlastec.2024.110637.
[34] R. Puerta *et al.*, “Analog Mobile Fronthaul for 6G and Beyond,” *Journal of Lightwave Technology*, vol. 42, no. 21, pp. 7458–7467, Nov. 2024, doi: 10.1109/JLT.2024.3435770.

[35] J. Yao, “Microwave Photonics,” *J. Lightwave Technol., JLT*, vol. 27, no. 3, pp. 314–335, Feb. 2009, Accessed: Jul. 17, 2026. [Online]. Available: https://opg.optica.org/jlt/abstract.cfm?uri=jlt-27-3-314
[36] P. Wang *et al.*, “Tunable band-pass microwave photonic filter with ultra-wide bandwidth and frequency tuning range based on cascaded tunable high-*Q* silicon micro-ring resonators,” *Photon. Res., PRJ*, vol. 14, no. 4, pp. 1517–1529, Apr. 2026, doi: 10.1364/PRJ.583471.
[37] Z. Tao *et al.*, “Ultrabroadband on-chip photonics for full-spectrum wireless communications,” *Nature*, vol. 645, no. 8079, pp. 80–87, Sep. 2025, doi: 10.1038/s41586-025-09451-8.
[38] T. J. Kippenberg, A. L. Gaeta, M. Lipson, and M. L. Gorodetsky, “Dissipative Kerr solitons in optical microresonators,” *Science*, vol. 361, no. 6402, p. eaan8083, Aug. 2018, doi: 10.1126/science.aan8083.
[39] B. Dong *et al.*, “Bidirectional W-Band Seamless Fiber-Wireless Integration System Enabled by Full-Photonic Up/Down Conversion at Optical Network Side,” *J. Lightwave Technol., JLT*, vol. 44, no. 1, pp. 77–86, Jan. 2026, Accessed: Jul. 17, 2026. [Online]. Available: https://opg.optica.org/jlt/abstract.cfm?uri=jlt-44-1-77
[40] K. Ye *et al.*, “Integrated RF Photonic Front-End Capable of Simultaneous Cascaded Functions”, doi: 10.1002/lpor.202401628.
[41] F. Eltes *et al.*, “A BaTiO3-Based Electro-Optic Pockels Modulator Monolithically Integrated on an Advanced Silicon Photonics Platform,” *Journal of Lightwave Technology*, vol. 37, no. 5, pp. 1456–1462, Mar. 2019, doi: 10.1109/JLT.2019.2893500.
[42] D. Liang and J. E. Bowers, “Recent Progress in Heterogeneous III-V-on-Silicon Photonic Integration,” *gxjzz*, vol. 2, no. 1, pp. 59–83, Mar. 2021, doi: 10.37188/lam.2021.005.
[43] Y. Hu *et al.*, “Integrated electro-optics on thin-film lithium niobate,” *Nat Rev Phys*, vol. 7, no. 5, pp. 237–254, May 2025, doi: 10.1038/s42254-025-00825-5.
[44] H. Ji, C. Sun, and W. Shieh, “Spectral Efficiency Comparison Between Analog and Digital RoF for Mobile Fronthaul Transmission Link,” *Journal of Lightwave Technology*, vol. 38, no. 20, pp. 5617–5623, Oct. 2020, doi: 10.1109/JLT.2020.3003123.
[45] M. U. Hadi, M. Y. Daha, S. K. O. Soman, and M. Ijaz, “Experimental Analysis of A-RoF Based Optical Communication System for 6G O-RAN Downlink,” in *2024 35th Irish Signals and Systems Conference (ISSC)*, Jun. 2024, pp. 1–6. doi: 10.1109/ISSC61953.2024.10603165.
[46] S. Rahman *et al.*, “Mitigation of Nonlinear Distortions for a 100 Gb/s Radio-Over-Fiber-Based WDM Network,” *Electronics*, vol. 9, no. 11, p. 1796, Nov. 2020, doi: 10.3390/electronics9111796.
[47] L. A. M. Pereira, L. L. Mendes, C. J. A. Bastos-Filho, and A. C. Sodré, “Novel Machine Learning Linearization Scheme for 6G A-RoF Systems,” *Journal of Lightwave Technology*, vol. 41, no. 23, pp. 7245–7252, Dec. 2023, doi: 10.1109/JLT.2023.3304281.
[48] M. U. Hadi, K. U. Danyaro, A. AlQushaibi, R. Qureshi, and T. Alam, “Digital Predistortion Based Experimental Evaluation of Optimized Recurrent Neural Network for 5G Analog Radio Over Fiber Links,” *IEEE Access*, vol. 12, pp. 19765–19777, 2024, doi: 10.1109/ACCESS.2024.3360298.
[49] M. U. Hadi, “Practical Demonstration of 5G NR Transport Over-Fiber System with Convolutional Neural Network,” *Telecom*, vol. 3, no. 1, pp. 103–117, Mar. 2022, doi: 10.3390/telecom3010006.
[50] C. Xie *et al.*, “Bidirectional WDM Multi-Nodes Analog Radio-Over-Fiber Mobile Fronthaul Link Enhanced by Photonic Integrated Devices,” *IEEE Photonics Journal*, vol. 14, no. 6, pp. 1–7, Dec. 2022, doi: 10.1109/JPHOT.2022.3220821.
[51] M. Usman Hadi, “Towards optimization of 5G NR transport over fiber links performance in 5G Multi-band Networks: An OMSA model approach,” *Optical Fiber Technology*, vol. 79, p. 103358, Sep. 2023, doi: 10.1016/j.yofte.2023.103358.
[52] H. Xu, A. Delmade, C. Browning, A. Atieh, Y. Yu, and L. P. Barry, “Demonstration of High Capacity Bidirectional A-RoF System Using Wavelength Reuse and Frequency Multiplexing,” *Journal of Lightwave Technology*, vol. 41, no. 8, pp. 2343–2350, Apr. 2023, doi: 10.1109/JLT.2022.3230741.
[53] C. L. M. P. Plazas, A. M. de Souza, D. R. Celino, and M. A. Romero, “Optimization of arrayed waveguide grating-filtering response for efficient analog radio-over-fiber fronthaul over a

wavelength-division multiplexing passive optical network," *Transactions on Emerging Telecommunications Technologies*, vol. 32, no. 1, p. e4113, 2021, doi: 10.1002/ett.4113.
[54] M. Niels *et al.*, "A high-speed heterogeneous lithium tantalate silicon photonics platform," *Nat. Photon.*, vol. 20, no. 2, pp. 225–231, Feb. 2026, doi: 10.1038/s41566-025-01832-9.